\documentclass[%
 reprint,
showpacs,
 amsmath,amssymb,
 aps,
prb,
]{revtex4-1}

\usepackage{graphicx}
\usepackage{dcolumn}
\usepackage{bm}

\begin{document}

\preprint{APS/123-QED}

\title{Current cross-correlation in the Anderson impurity model with exchange interaction}

\author{Rui Sakano$^{1,2}$}
\email{sakano@issp.u-tokyo.ac.jp}
\author{Akira Oguri$^3$}%
\author{Yunori Nishikawa$^3$}
\author{Eisuke Abe$^4$}
\affiliation{%
 $^1$Institute for solid state physics, the university of Tokyo, 5-1-5 Kashiwanoha, Kashiwa, Chiba, 277-8581 Japan \\
 $^2$Laboratoire de Physique des Solides, Universit\'e Paris-Sud, CNRS, UMR 8502, F-91405 Orsay Cedex, France \\
 $^3$Department of physics, Osaka city university, 3-3-138 Sugimoto Sumiyoshi-ku, Osaka-shi, 558-8585 Japan \\
 $^4$Spintronics Research Center, Keio University, 3-14-1 Hiyoshi, Kohoku-ku, Yokohama 223-8522, Japan
}%




\date{\today}

\begin{abstract}
We study spin-entanglement of the quasiparticles of the local Fermi liquid excited
in nonlinear current through a quantum dot
described by the Anderson impurity model
with two degenerate orbitals coupled to each other via
an exchange interaction. 
Applying the renormalized perturbation theory,
we obtain
the precise form of the cumulant generating function
and cross-correlations
for the currents with spin angled to arbitrary directions,
up to third order in the applied bias voltage. 
It is found that the exchange interaction gives rise to
spin-angle dependency in the cross-correlation
between the currents through the two different orbitals,
and also brings
an intrinsic cross-correlation of currents with three different angular momenta.
\end{abstract}

\maketitle


\section{Introduction}

In dilute magnetic alloys,
a magnetic moment in an impurity
and that of surrounding conduction electrons in the host metal
form a singlet ground state at low temperatures.  
This phenomenon, known as the Kondo effect, 
has been intensively studied as a central issue
of the condensed matter physics from the impurity problem to the heavy Fermions
since the discovery of the mechanism.
\cite{HewsonBook}
Recently, development in modern condensed matter systems
such as semiconductor quantum dots, ultracold atoms, and condensed quarks,
has stimulated further research to explore new aspects of the Kondo effect.
\cite{Nature391-156,Wiel2105,1367-2630-18-7-075012,PhysRevD.92.065003,PhysRevD.94.074013} 
In quantum dots, the Kondo effect in the nonequilibrium steady state
beyond the linear response regime
has been achieved by applying small bias-voltages,
which has been shedding light on a new aspect of
the local Fermi liquid.
\cite{PhysRevB.64.153305,PhysRevLett.100.246601,PhysRevB.84.245316,
PhysRevLett.97.016602,PhysRevLett.97.086601,PhysRevLett.100.036603,PhysRevLett.100.036604,PhysRevB.80.155322,doi:10.1143/JPSJ.79.044714,PhysRevLett.108.266401,
PhysRevB.77.241303,NaturePhys5.208,PhysRevLett.106.176601,NatPhys.12.230,PhysRevB.83.075440,PhysRevB.83.241301,PhysRevLett.118.196803}
The local Fermi liquid is an extension of Landau's Fermi liquid theory
to the Kondo impurity at low energies. 
It is essentially accounted for
by free quasiparticles and renormalized interactions.
\cite{Nozieres1974,doi:10.1143/PTP.53.970,PhysRevB.64.153305}
In electric currents through quantum dots in the Kondo regime,
the renormalized interaction excites pairs of quasiparticles
that give rise to backscattering currents
with an effective charge of $2e$.
\cite{PhysRevLett.97.016602,PhysRevLett.97.086601,PhysRevLett.100.036603,PhysRevLett.100.036604,PhysRevB.80.155322,doi:10.1143/JPSJ.79.044714,PhysRevLett.108.266401}
This doubly-charged state enhances fluctuation of the electric current through the quantum dot,
which has been observed
as enhancement of the shot noise or the Fano factor.
\cite{PhysRevB.77.241303,NaturePhys5.208,PhysRevLett.106.176601,NatPhys.12.230,PhysRevB.83.075440,PhysRevB.83.241301,PhysRevLett.118.196803}
The shot noise in the Kondo dots has elucidated that
interacting quasiparticles form charge pairs
in the nonlinear current 
driven by the applied bias voltage.
A question that remains to be answered is if the spins of the quasiparticle pair are entangled.

The purpose of this paper is to explore the nature
of the spin entanglement of pairs of the quasiparticles
excited by the renormalized interaction of the local Fermi liquid
in the current.
We emphasize that our focus is the spin entanglement between quasiparticles.
So far,
a lot of works on 
the spin entanglement
between the impurity and the conduction electrons,
such as entropy of the impurity and the Kondo cloud, have been done.
\cite{Affleck2010}
Some theoretical works on cross-correlations between currents 
with different channels 
in the SU($N$) Kondo quantum dot have also been done.
\cite{PhysRevB.76.241307,PhysRevB.75.235105,PhysRevB.81.241305,PhysRevB.83.241301}
The cross-correlation arises in the nonlinear current,
due to the excited charge pair,
and its bias-voltage dependence is 
universally scaled by the Kondo temperature.
However, the cross-correlation is independent of the spin angle of the currents
in the SU($N$) Kondo regime, 
where the renormalized interactions are isotropic
for spins of the interacting quasiparticles.
Therefore, the quasiparticle's entanglement may sensitively 
depend on the renormalized interaction in the case where it 
acquires spin-dependent components.

In this paper,
to assess the spin entanglement of the quasiparticle pairs,
we introduce the Anderson impurity model 
with degenerate orbitals
which couple each other via an exchange interaction.
Particularly, we investigate the inte-orbital cross-correlations 
of two currents with twisted spin-angles,
and also
cross-correlations between three currents through different channels.
To  this end,
we make use of the full counting statistics to calculate the current cross-correlations,
which can give us all the necessary components of the 
current correlations systematically.
\cite{PROP:PROP200610305,RevModPhys.81.1665}
The renormalized perturbation theory is also 
 employed to precisely account for effects of
electron correlation in
low bias voltage steady state.
\cite{PhysRevLett.70.4007,0953-8984-13-44-314,doi:10.1143/JPSJ.74.110,PhysRevB.82.115123}

This paper is organized as follows.
In the next section, we introduce
an orbital-degenerate Anderson impurity model with an exchange interaction.
The renormalized perturbation theory is introduced
to precisely treat electron correlations of the local Fermi liquid
in the nonequlibrium steady state at low bias voltages.
The full counting statics is also introduced
to calculate current correlations. 
In Sec. \ref{sec:result}, 
we show  
our numerical renormalization group results to the
interaction-dependence of the current cross-correlations, 
and discuss 
the entangled states
in the nonlinear current
in the Fermi liquid regime.
Finally,
we give a brief summary in Sec. \ref{sec:summary}.

\section{Model and formulation}

\subsection{Anderson Impurity model}
\begin{figure}[tb]
	\includegraphics[width=6cm]{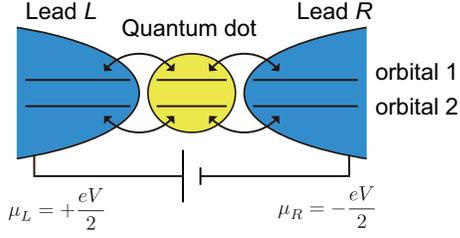}
	\caption{\label{fig:mdl}
	A schematic of a quantum dot connected to left and right lead electrodes with orbital degrees of freedom.
	Bias voltage $eV$ is applied between the two lead electrodes.
	In electron tunneling between the dot and the leads, the spin and orbital degrees of freedom are assumed to be conserved.}
\end{figure}

Let us consider a quantum dot with two degenerate orbitals coupled by an exchange interaction (see Fig. \ref{fig:mdl}),
described by an Anderson impurity model:
\begin{eqnarray}
	{\cal H}_{\rm A}^{} &=& {\cal H}_{0}^{} + {\cal H}_{\rm T}^{}  + {\cal H}_{\rm I}^{} 
	\, ,
	\label{eq:ImpurityAndersonModel}
\end{eqnarray}
where
\begin{eqnarray}
	{\cal H}_{0}^{} &=&
	\sum_{\alpha m \sigma} \int_{-D}^{D} d\varepsilon \, \varepsilon \,
	c_{\varepsilon \alpha m \sigma}^{\dagger} c_{\varepsilon \alpha m \sigma}^{}
	+ \sum_{m \sigma} \epsilon_{\rm d}^{} d_{m\sigma}^{\dagger}  d_{m\sigma}^{} \, , \\
{\cal H}_{\rm T}^{} &=&
\sum_{\alpha m\sigma} \left[ v_{\alpha}^{} d_{m\sigma}^{\dagger} \psi_{\alpha m \sigma}^{}
+ v_{\alpha}^{*} \psi_{\alpha m \sigma}^{\dagger} d_{m\sigma}^{} 
\right] \, , \\
{\cal H}_{\rm I}^{} &=&
U \sum_{m} \hat{n}_{{\rm d}m\uparrow}^{} \hat{n}_{{\rm d}m\downarrow}^{}
+ W \hat{n}_{{\rm d}1}^{} \hat{n}_{{\rm d}2}^{}
+ 2J \hat{\bm S}_{{\rm d}1}^{} \cdot \hat{\bm S}_{{\rm d}2}^{} \, .
\end{eqnarray}
The first term ${\cal H}_0^{}$ represents electrons
in the two lead electrodes and the quantum dot.
The operator $c_{\alpha\epsilon m\sigma}^{}$ annihilates an electron
with spin $\sigma = \uparrow, \downarrow$, orbital $m=1,2$ and energy $\varepsilon$
in the conduction band of the left and right electric leads $\alpha = L, R$.
The operator $d_{m\sigma}^{}$ annihilates an electron with spin $\sigma$ and orbital $m$
in the dot level $\epsilon_{\rm d}^{}$.
The second term ${\cal H}_{\rm T}^{}$ represents
the electron tunneling between the leads and the dot.
The leads and the dot are connected by tunneling matrix element $v_{\alpha}^{}$
through
$\psi_{\alpha m \sigma}^{} := \int_{-D}^{D} d\varepsilon \sqrt{\rho_{\rm c}^{}} c_{\varepsilon \alpha m \sigma}$
where $D$ is the half width of the conduction band
 and $\rho_{\rm c}^{}=\frac{1}{2D}$ is the density of state for the conduction electrons.
The electron tunneling leads to an intrinsic linewidth of the dot level given by
 $\Gamma = \frac{1}{2} \left( \Gamma_L^{} + \Gamma_R^{} \right)$
 with $\Gamma_{\alpha}^{} := 2 \pi \rho_{\rm c}^{} \left| v_{\alpha}^{} \right|^2$.
The last term ${\cal H}_{\rm I}^{}$ represents the interaction between the electrons in the dot, where
$U$ and $W$ are the intra- and interorbital Coulomb interaction, respectively, and
$J$ is the exchange interaction.
The number operators of the electrons in the dot are defined by
$\hat{n}_{{\rm d}m\sigma}^{} := d_{m\sigma}^{\dagger} d_{m\sigma}^{}$ and
$\hat{n}_{{\rm d}m}^{} := \sum_{\sigma} \hat{n}_{{\rm d}m\sigma}^{}$,
and the total spin operator for the electrons in the dot in channel $m$ is defined by
$\hat{\bm S}_{{\rm d} m}^{} := \sum_{\sigma \sigma'}d_{m\sigma}^{\dagger} {\bm \sigma}_{\sigma \sigma'} d_{m\sigma'}^{}$,
where ${\bm \sigma}$ is the Pauli matrix.

In calculation of current correlations, we assume
particle-hole symmetric $\epsilon_{\rm d}^{} = - \frac{U}{2} -W$, symmetric connection $\Gamma_L^{}= \Gamma_R^{}$,
and the absolute zero temperature $T=0$
to eliminate thermal and partition noise
and emphasize spin-entanglement due to the exchange interaction $J$.
The bias voltage $eV$ is symmetrically applied
 between the left and right leads to induce electric current:
The chemical potential of the left and right  leads are
$\mu_{L}^{} = +\frac{1}{2}eV$ and
$\mu_R^{}=- \frac{1}{2}eV$, respectively. 
Without loss of generality,
positive bias voltage $eV>0$ can be taken.
We also use the natural unit $\hbar=k_{\rm B}^{}=e=1$.

We investigate spin entanglement of interacting quasiparticle pairs
emerging in nonlinear currents through the two orbitals,
exploiting cross-correlations for current with two twisted spin angles.
The operator of the electric current with spin angled to the $\theta$ direction, from the lead $\alpha$ to the dot
can be defined by  
\begin{eqnarray}
	I_{\alpha m \theta}^{}
	= -i \left( v_{\alpha}^{} d_{m\theta}^{\dagger} \psi_{\alpha m\theta}^{}
	- v_{\alpha}^{*} \psi_{\alpha m\theta}^{\dagger} d_{m\sigma}^{} \right)\, .
\end{eqnarray}
Here the operators for the electrons with spin angled to the $\theta$ direction can be defined
by a rotational transformation without loss of generality:
\begin{eqnarray}
	\begin{pmatrix}
		d_{m\theta}^{} \\
		d_{m\, \theta+\pi}^{}
	\end{pmatrix}
	&:=&
	\begin{pmatrix}
		\cos\frac{\theta}{2} & - \sin \frac{\theta}{2} \\
		\sin \frac{\theta}{2} & \cos\frac{\theta}{2}
	\end{pmatrix}
	\begin{pmatrix}
		d_{m\uparrow}^{} \\
		d_{m\downarrow}^{}
	\end{pmatrix}
	\, , \\
	\begin{pmatrix}
		\psi_{\alpha m\theta}^{} \\
		\psi_{\alpha m \, \theta+\pi}^{}
	\end{pmatrix}
	&:=&
	\begin{pmatrix}
		\cos\frac{\theta}{2} & - \sin \frac{\theta}{2} \\
		\sin \frac{\theta}{2} & \cos\frac{\theta}{2}
	\end{pmatrix}
	\begin{pmatrix}
		\psi_{\alpha m\uparrow}^{} \\
		\psi_{\alpha m\downarrow}^{}
	\end{pmatrix}
	\, .
\end{eqnarray}

We note that our analysis can be applied
to not only single quantum dots with orbital degeneracy
\cite{PhysRevLett.93.017205,nature03422}
but also double quantum dots with two current channels.
\cite{nphys2844}

\subsection{Renormalized perturbation theory}
To derive the precise form of current cross-correlations 
under the electron correlations
of the Anderson impurity model at low energies,
we make use of the renormalized perturbation theory.

The renormalized perturbation theory is an idea to reorganize the series of perturbation expansion,
which is very useful
for systems where the renormalization effect strongly acts such as Kondo impurities.
The renormalized perturbation theory for the Anderson impurity model links
the {\it microscopic} local Fermi liquid theory where perturbation expansion is done in powers of bare interactions
\cite{doi:10.1143/PTP.53.970,doi:10.1143/PTP.55.67}
to {\it phenomenological} local Fermi liquid theory
where perturbation expansion is done in powers of the renormalized interactions.
\cite{Nozieres1974}
Thus, the theory tells us the precise way to calculate currents and current-correlations at low energies,
by perturbation expansion in the renormalized interactions,
and brings an intuitive understanding of 
the current due to low-lying excited states
in the quasiparticle picture.

In this subsection,
we illustrate the renormalized perturbation theory
for the Anderson impurity model with degenerate orbitals
given by Eq. (\ref{eq:ImpurityAndersonModel}).
\cite{PhysRevLett.70.4007,0953-8984-13-44-314,PhysRevB.82.115123}
Then, we apply this theory to
calculate current cross-correlations up to third order in bias-voltage $V^3$.
The basic idea is as the following.
In the local Fermi liquid region, perturbation expansion
in the interactions $U, W$, and $J$
for all orders gives the exact result at low energies.
However, it is very difficult, except for some special cases,
to calculate all series in the perturbation expansion.
Here, employing the idea of the renormalized perturbation theory,
we reorganize the perturbation expansion
and effectively carry out all-order calculation at low energies.

\begin{figure}[tb]
	\includegraphics[width=6cm]{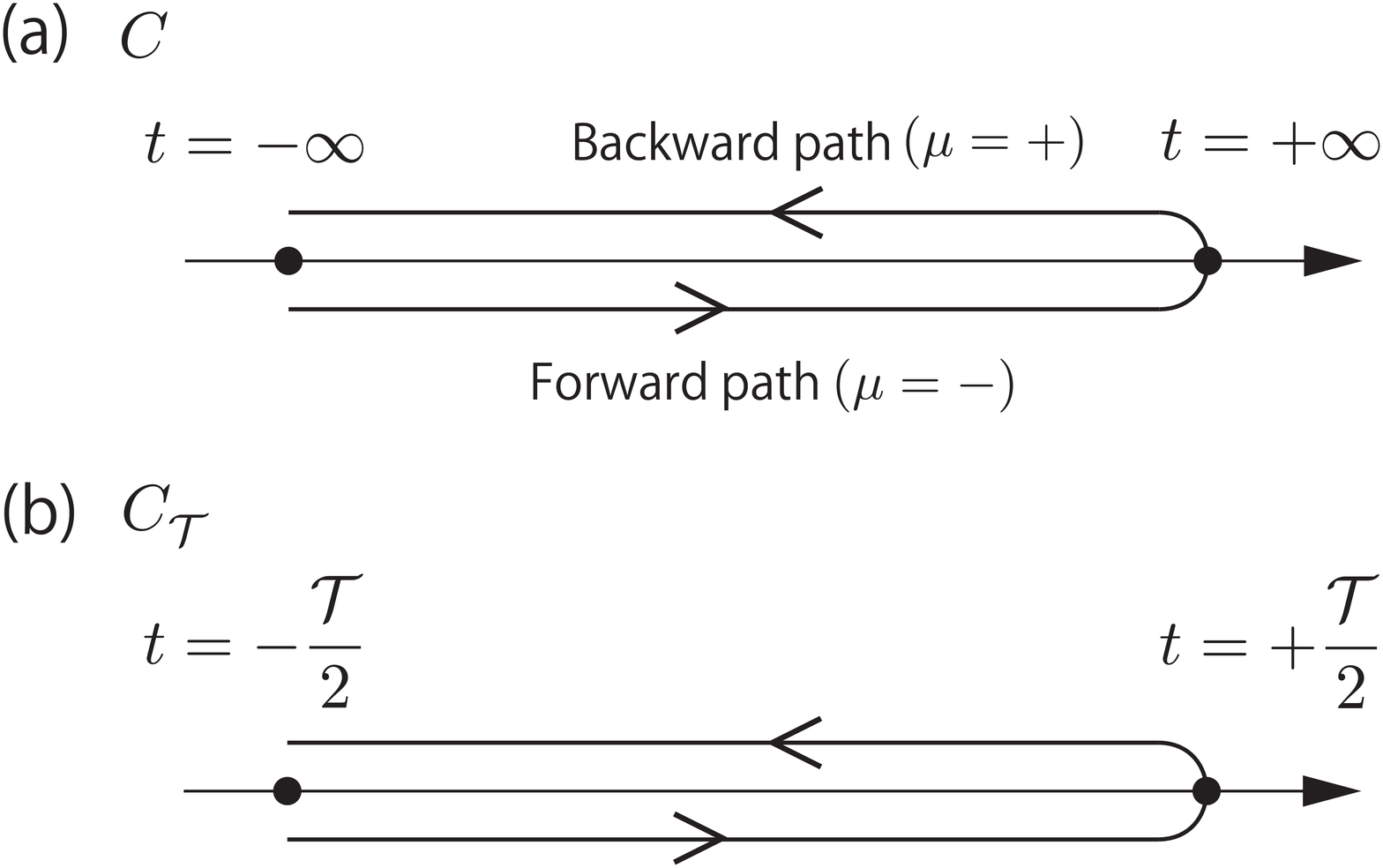}
	\caption{\label{fig:KC}
	The Keldysh contour
	(a) $C$ between $t=-\infty$ and $t=+\infty$, and
	(b) $C_{\cal T}^{}$ between $t=-\frac{\cal T}{2}$ and $t=+\frac{\cal T}{2}$.
	The indices $\mu= -$ and $+$ specify the forward and backward paths, respectively.}
\end{figure}

Let us start with the partition function
for the Anderson impurity model given by Eq. (\ref{eq:ImpurityAndersonModel}),
which can be expressed
as a functional integral over time-dependent Grassmann variables along the Keldysh contour $C$
[see Fig. \ref{fig:KC} (a)],
\begin{eqnarray}
	{\cal Z}=
	\int \prod_{\alpha m \sigma}
	{\cal D}\left( \bar{d}_{m\sigma}^{} \right)  {\cal D}\left( d_{m\sigma}^{} \right)  
	{\cal D}\left( \bar{c}_{\varepsilon \alpha m\sigma}^{} \right)  {\cal D}\left( c_{\varepsilon \alpha m\sigma}^{} \right) 
	e^{i{\cal S}}
	\, ,
	\nonumber \\
\end{eqnarray}
where the action is given by
\begin{eqnarray}
{\cal S} =
\sum_{\mu} \int_{-\infty}^{\infty} dt \, \left( \sigma_3^{}\right )^{\mu\mu} {\cal L}_{\rm A}^{\mu} ( t) .
\end{eqnarray}
with
\begin{eqnarray}
	\sigma_3^{} =
	\begin{pmatrix}
	1 & 0 \\
	0 & -1 
	\end{pmatrix}
	.
\end{eqnarray}
The superscripts $\mu= -$ and $+$ label the forward and backward paths of the Keldysh contour, respectively,
as shown in Fig. \ref{fig:KC} (a).
The integral along the Keldysh contour allows us
to calculate expectation values in nonequilibrium states.
Our Anderson impurity model in the Lagrangean form is given by
\begin{eqnarray}
	{\cal L}_{\rm A}^{\mu} ( t) =
	{\cal L}_{0}^{\mu} ( t) + {\cal L}_{\rm T}^{\mu} ( t ) + {\cal L}_{\rm I}^{\mu} ( t)
	\, ,
\end{eqnarray}
with
\begin{eqnarray}
	{\cal L}_{0}^{\mu} ( t ) &=&
	\sum_{\alpha m \sigma} \int_{-D}^{D} d\varepsilon \,
	\bar{c}_{\varepsilon \alpha m \sigma}^{\mu} ( t )
	\left( i\frac{ \partial}{\partial t}-\varepsilon \right)
	c_{ \varepsilon \alpha m \sigma}^{\mu} ( t)
	\nonumber \\
	&& \qquad
	+ \sum_{m \sigma}
	\bar{d}_{m\sigma}^{\mu} ( t )
	\left( i \frac{\partial}{\partial t} - \epsilon_{\rm d}^{} \right)
	d_{m\sigma}^{\mu} ( t) \, , \\
	{\cal L}_{\rm T}^{\mu} ( t) &=&
	\sum_{\alpha m\sigma}
	\left[
	v_{\alpha}^{} \bar{d}_{m\sigma}^{\mu} ( t) \, \psi_{\alpha m \sigma}^{\mu} ( t )
	\right.
	\nonumber\\
&& \qquad \left.
+ v_{\alpha}^{*} \bar{\psi}_{\alpha m \sigma}^{\mu} ( t ) \, d_{m\sigma}^{\mu} ( t)
\right] \, , \\
	{\cal L}_{\rm I}^{\mu} \left( t \right) &=&
	U \sum_{m} n_{{\rm d}m\uparrow}^{\mu} ( t) \, n_{{\rm d}m\downarrow}^{\mu} ( t) \nonumber \\
	&& + W n_{{\rm d}1}^{\mu} ( t ) \, n_{{\rm d}2}^{\mu} ( t)
	+ 2J {\bm S}_{{\rm d}1}^{\mu} ( t) \cdot {\bm S}_{{\rm d}2}^{\mu} ( t) \, .
	\label{eq:L-q-int}
\end{eqnarray}
Here, the Grassmann numbers are defined by
\begin{eqnarray}
n_{{\rm d}m\sigma}^{\mu} (t)&:=&
\bar{d}_{m\sigma}^{\mu} (t) d_{m\sigma}^{\mu} (t) \, , \\
n_{{\rm d}m}^{\mu} (t)&:=& \sum_{sigma} n_{{\rm d}m\sigma}^{\mu} (t) \, , \\
{\bm S}_{{\rm d} m}^{\mu} (t) &:=&
\sum_{\sigma \sigma'} \bar{d}_{m\sigma}^{\mu} (t) {\bm \sigma}_{\sigma \sigma'} d_{m\sigma'}^{\mu} (t) \, , \\
\psi_{\alpha m \sigma}^{\mu} (t) &:=&
\int_{-D}^{D} d\varepsilon \sqrt{\rho_{\rm c}^{}} c_{\varepsilon \alpha m \sigma}^{\mu} (t) , \\
\bar{\psi}_{\alpha m \sigma}^{\mu}  (t) &:=&
\int_{-D}^{D} d\varepsilon \sqrt{\rho_{\rm c}^{}} \bar{c}_{\varepsilon \alpha m \sigma}^{\mu} (t) .
\end{eqnarray}

On introducing a self-energy $\Sigma_{{\rm d}m\sigma}^{\rm r} ( \omega)$
due to the interactions, $U,W$, and $J$,
the full retarded Green's function for the electrons in the dot can be written in the form
\begin{eqnarray}
	G_{{\rm d}m\sigma}^{\rm r} ( \omega ) = \frac{1}{\omega - \epsilon_{\rm d}^{} + i \Gamma - \Sigma_{{\rm d}m\sigma}^{\rm r} ( \omega)}
	\, . \label{eq:FllGrnFnctn}
\end{eqnarray}
We reorganize the perturbation expansion in the bare interactions $U,W$, and $J$
 to a form appropriate for the low energies.
The first step is to write the self-energy in the form
\begin{eqnarray}
	\Sigma_{{\rm d}m\sigma}^{\rm r} ( \omega ) =
	\Sigma_{{\rm d}m\sigma}^{\rm r} ( 0 )
	+ \omega  \,  \Sigma_{{\rm d}m\sigma}^{{\rm r}\, \prime} ( 0 )
	 + \Sigma_{{\rm d} m\sigma}^{\rm rem} ( \omega ) \, ,
\end{eqnarray}
which defines the remainder self-energy
$\Sigma_{{\rm d}m\sigma}^{\rm rem} ( \omega )$.
Here, $\Sigma_{{\rm d}m\sigma}^{{\rm r}\, \prime} ( \omega) := \frac{\partial  \Sigma_{{\rm d}m\sigma}^{\rm r} ( \omega)}{\partial \omega}$.
Substituting the self-energy in this form into the full Green's function given in Eq. (\ref{eq:FllGrnFnctn}),
Green's function of the quasiparticles $\widetilde{G}_{{\rm d}m\sigma}^{\rm r} \left( \omega \right)$
takes the same form with
a renormalized energy level of the localized state 
$\tilde{\epsilon}_{\rm d}^{}:= z\left[ \epsilon_{\rm d}^{} + \Sigma_{{\rm d}m\sigma}^{\rm r} ( 0)\right]$,
level width
$\widetilde{\Gamma} := z \Gamma$,
and self-energy
$\widetilde{\Sigma}_{{\rm d}m\sigma}^{\rm r} ( \omega)
:= z \Sigma_{{\rm d}m\sigma}^{\rm rem} ( \omega )$:
\begin{eqnarray}
	G_{{\rm d}m\sigma}^{\rm r} ( \omega )
	&=& \frac{z}{\omega - \tilde{\epsilon}_{\rm d}^{} + i \widetilde{\Gamma}
	- \widetilde{\Sigma}_{{\rm d}m\sigma}^{\rm r} 	( \omega )} \\
	&=& z \widetilde{G}_{{\rm d}m\sigma}^{\rm r} ( \omega )
	\, .
\end{eqnarray}
Here, the wave function renormalization factor is defined by
$z:=\left[ 1- \Sigma_{{\rm d}m\sigma}^{{\rm r}\, \prime} ( 0) \right]^{-1}$.
The renormalized linewidth $\widetilde{\Gamma}$ corresponds to the characteristic energy scale,
namely, the Kondo temperature as
$T_{\rm K} = \pi \widetilde{\Gamma}/4$.
We note that $\Sigma_{{\rm d}m\sigma}^{\rm r} ( 0)$
and $\Sigma_{{\rm d}m\sigma}^{{\rm r}\, \prime} ( 0 )$
are to be evaluated at $\omega=0, T=0$, and $V = 0$.
The overall factor $z$ in Green's function
is removed by rescaling the Grassmann numbers of the dot state as, 
$d_{m\sigma}= \sqrt{z} \tilde{d}_{m\sigma}$
and $\bar{d}_{m\sigma} = \sqrt{z} \tilde{\bar{d}}_{m\sigma}$.

The last parameters specifying the renormalized theory are the renormalized interactions
$\widetilde{U}, \widetilde{W}$, and $\widetilde{J}$.
These quantities are derived from the four-point vertex function
$\Gamma_{m_3 \sigma_3 ; m_4 \sigma_4 }^{m_1 \sigma_1 ; m_2 \sigma_2} ( \omega_1^{}, \omega_2^{}, \omega_3^{}, \omega_4^{})$,
which is derived from the time-ordering two-particles Green's function of the dot electrons
at $V=0$ and $T=0$.
The renormalized four-vertex is defined by
\begin{eqnarray}
&& \widetilde{\Gamma}_{m_3 \sigma_3 ; m_4 \sigma_4 }^{m_1 \sigma_1 ; m_2 \sigma_2} \left( \omega_1^{}, \omega_2^{}, \omega_3^{}, \omega_4^{} \right)
\nonumber \\
&& \qquad  := z^2 \Gamma_{m_3 \sigma_3 ; m_4 \sigma_4 }^{m_1 \sigma_1 ; m_2 \sigma_2} \left( \omega_1^{}, \omega_2^{}, \omega_3^{}, \omega_4^{} \right)
\, ,
\end{eqnarray}
through
the rescaling of four fermion fields for the dot site.
The renormalized interactions are then defined by the value of the four-vertex
at $\omega_1^{}=\omega_2^{}=\omega_3^{}=\omega_4^{}=0$:
\begin{eqnarray}
	&& z^2 \Gamma_{m_3 \sigma_3 ; m_4 \sigma_4 }^{m_1 \sigma_1 ; m_2 \sigma_2} \left( 0, 0, 0, 0 \right) \nonumber \\
	&& \qquad =: \left[ \left( \widetilde{U} +  \widetilde{J} \right) \delta_{m_2}^{m_1}
	+\left( \widetilde{W} - \frac{\widetilde{J}}{2} \right) \left( 1 - \delta_{m_2}^{m_1} \right) \right]
	\nonumber \\
	&& \qquad \qquad  \qquad \times
	\left( \delta_{m_4}^{m_1} \delta_{m_3}^{m_2} \delta_{\sigma_4}^{\sigma_1} \delta_{\sigma_3}^{\sigma_2} 
	- \delta_{m_3}^{m_1} \delta_{m_4}^{m_2} \delta_{\sigma_3}^{\sigma_1} \delta_{\sigma_4}^{\sigma_2} 
	\right) \nonumber \\ 
	&& \qquad \qquad - \widetilde{J}
	\left( \delta_{m_3}^{m_1} \delta_{m_4}^{m_2} \delta_{\sigma_4}^{\sigma_1} \delta_{\sigma_3}^{\sigma_2} 
	- \delta_{m_4}^{m_1} \delta_{m_3}^{m_2}  \delta_{\sigma_3}^{\sigma_1} \delta_{\sigma_4}^{\sigma_2} 
	\right)
	\, ,
\end{eqnarray}
where
$\delta_{m_2}^{m_1} $ is the Kronecker's delta.

The quasiparticle's Lagrangean $\widetilde{\cal L}_{\rm qp}^{\mu} ( t)$
to describe properties at low energies
is obtained
by replacing the parameters of the dot state of the Anderson impurity model
given by Eq. (\ref{eq:ImpurityAndersonModel}) to the renormalized ones:
\begin{eqnarray}
	\widetilde{\cal L}_{\rm qp}^{\mu} ( t) &=&
	\widetilde{\cal L}_{0}^{\mu} ( t)
	+ \widetilde{\cal L}_{\rm T}^{\mu} ( t)
	+ \widetilde{\cal L}_{\rm I}^{\mu} ( t )
	\, ,
\end{eqnarray}
where
\begin{eqnarray}
	\widetilde{\cal L}_{0}^{\mu} \left( t \right) &=&
	\sum_{\alpha m \sigma} \int_{-D}^{D} d\varepsilon \,
	\bar{c}_{\varepsilon \alpha m \sigma}^{\mu}
	\left( t \right) \left( i\frac{ \partial}{\partial t}-\varepsilon \right)
	c_{ \varepsilon \alpha m \sigma}^{\mu} \left( t \right)\nonumber \\
	&& \qquad
	+ \sum_{m \sigma}
	\tilde{\bar{d}}_{m\sigma}^{\mu} \left( t \right)
	\left( i \frac{\partial}{\partial t} - \tilde{\epsilon}_{\rm d}^{} \right)
	\tilde{d}_{m\sigma}^{\mu} \left( t \right) \, , \\
	\widetilde{\cal L}_{\rm T}^{\mu} \left( t \right) &=&
	\sum_{\alpha m\sigma}
	\left[
	\tilde{v}_{\alpha}^{} \tilde{\bar{d}}_{m\sigma}^{\mu}\left( t \right) \psi_{\alpha m \sigma}^{\mu}\left( t \right)
	\right.
	\nonumber\\
	&& \qquad \left.
	+ \tilde{v}_{\alpha}^{*} \bar{\psi}_{\alpha m \sigma}^{\mu} \left( t \right) \tilde{d}_{m\sigma}^{\mu} \left( t \right)
	\right] \, , \\
	\widetilde{\cal L}_{\rm I}^{\mu} \left( t \right) &=&
	\widetilde{U} \sum_{m} \tilde{n}_{{\rm d}m\uparrow}^{\mu}\left( t \right) \tilde{n}_{{\rm d}m\downarrow}^{\mu}\left( t \right)
	\nonumber \\
	&&+ \widetilde{W}
	\tilde{n}_{{\rm d}1}^{\mu}\left( t \right) \tilde{n}_{{\rm d}2}^{\mu} \left( t \right)
	+ 2\widetilde{J} \widetilde{\bm S}_{{\rm d}1}^{\mu}\left( t \right) \cdot \widetilde{\bm S}_{{\rm d}2}^{\mu}\left( t \right)
	\, ,
	\label{eq:lagrangean4qpInt}
\end{eqnarray}
with
\begin{eqnarray}
	\tilde{v}_{\alpha}^{} &:=& \sqrt{z} v_{\alpha}^{}
	\, , \\
	\tilde{n}_{{\rm d}m\sigma}^{\mu} (t)&:=&
	\tilde{\bar{d}}_{m\sigma}^{\mu} (t) \tilde{d}_{m\sigma}^{\mu} (t)
	\, , \\
	\tilde{n}_{{\rm d}m}^{\mu} (t)&:=&
	\sum_{\sigma} \tilde{n}_{{\rm d}m\sigma}^{\mu} (t)
	\, , \\
	\widetilde{\bm S}_{{\rm d} m}^{\mu} (t) &:=&
	\sum_{\sigma \sigma'} \tilde{\bar{d}}_{m\sigma}^{\mu} (t) {\bm \sigma}_{\sigma \sigma'} \tilde{d}_{m\sigma'}^{\mu} (t)
	\, .
\end{eqnarray}
As a part
of the interaction effects are taken into account {\it ab initio} in the quasiparticle's Lagrangean,
compensating terms have to be introduced to avoid overcounting in the perturbation expansion.
Then, the total Lagrangean has to be satisfied with
\begin{eqnarray}
	{\cal L}_{\rm A}^{\mu} ( t ) = \widetilde{\cal L}_{\rm qp}^{\mu} ( t )  + {\cal L}_{\rm CT}^{\mu} ( t) .
	\label{eq:Idntty-Lgrngns}
\end{eqnarray}
Therefore, the specific form of the counter-term Lagrangean is given by
\begin{eqnarray}
	{\cal L}_{\rm CT}^{\mu} (t)
	&=& \sum_{m \sigma} \tilde{\bar{d}}_{m\sigma}^{\mu} ( t )
	\left( \xi_1^{} i\frac{\partial}{\partial t} + \xi_2^{} \right)
	\tilde{d}_{m\sigma}^{\mu} ( t ) \nonumber \\
	&& + \xi_{3}^{U} \sum_{m} \tilde{n}_{{\rm d}m\uparrow}^{\mu} ( t ) \, \tilde{n}_{{\rm d}m\downarrow}^{\mu} ( t)
	+ \xi_{3}^{W}  \tilde{n}_{{\rm d}1}^{\mu} ( t) \, \tilde{n}_{{\rm d}2}^{\mu} ( t) \nonumber \\
	&& + 2 \xi_{3}^{J} \widetilde{\bm S}_{{\rm d}1}^{\mu} ( t ) \cdot \widetilde{\bm S}_{{\rm d}2}^{\mu} ( t)
	\, .
	\label{eq:Lagrangean4CT}
\end{eqnarray}
In the reorganized perturbation theory,
the action can be written in the terms of the quasiparticles.
After formally integrating over the Grassmann numbers for the conduction electrons,
the reduced action is given by
\begin{eqnarray}
{\cal S} &=&
\sum_{\mu} \int_{-\infty}^{\infty} dt \left( \sigma_3^{} \right)_{}^{\mu\mu} 
\left[
\widetilde{\cal L}_{\rm qp}^{\mu} \left( t \right) + {\cal L}_{\rm CT}^{\mu}( t )
\right] \\
&=&
\int_{-\infty}^{\infty} dt \int_{-\infty}^{\infty} dt' \sum_{m\sigma}
\tilde{\bar{\bm d}}_{m\sigma}^{} ( t )
\left[ \widetilde{\bm g}_{{d}m\sigma}^{} ( t-t' ) \right]^{-1}
\tilde{\bm d}_{m\sigma}^{} ( t' )
\nonumber \\
&& \qquad + \sum_{\mu} \left( \sigma_3^{} \right)_{}^{\mu\mu}  \int_{-\infty}^{\infty} dt\,
\left[ 
\widetilde{\cal L}_{\rm I}^{\mu} ( t )
+ {\cal L}_{\rm CT}^{\mu} ( t )
\right]
\label{eq:ActionInQP}
\end{eqnarray}
with
\begin{eqnarray}
&& \tilde{\bar{\bm d}}^{} ( t ) =
 \begin{pmatrix}
\tilde{d}_{{\rm d}m\sigma}^{-} ( t ) &
 \tilde{d}_{{\rm d}m\sigma}^{+} ( t )
\end{pmatrix}
, \ 
\tilde{\bm d} ( t ) =
 \begin{pmatrix}
\tilde{d}_{{\rm d}m\sigma}^{-} ( t ) \\
 \tilde{d}_{{\rm d}m\sigma}^{+} ( t )
\end{pmatrix}
, \\
&&
\widetilde{\bm g}_{{\rm d}m\sigma}^{} ( t )=
 \begin{pmatrix}
\tilde{g}_{{\rm d}m\sigma}^{--} ( t ) & \tilde{g}_{{\rm d}m\sigma}^{-+} ( t ) \\
\tilde{g}_{{\rm d}m\sigma}^{+-} ( t ) & \tilde{g}_{{\rm d}m\sigma}^{++} ( t )
\end{pmatrix}
\, .
\end{eqnarray}
Here, $\widetilde{\bm g}_{{\rm d}m\sigma}^{} ( t )$ is Green's function of the free quasiparticle (see Appendix \ref{secA:GF4FQ}).
The coefficients of the counter terms,
$\xi_{1}^{}, \xi_2^{}, \xi_3^{U}, \xi_3^{W}$, and $\xi_3^J$, are expressed
in powers of the renormalized interactions
$\widetilde{U}, \widetilde{W}$ and $\widetilde{J}$,
which are determined
by the renormalized condition for the renormalized self-energy given
$\widetilde{\Sigma}_{{\rm d}m\sigma}^{\rm r} ( 0 ) = 0$
and
$\left. \frac{ \partial \widetilde{\Sigma}_{{\rm d}m\sigma}^{\rm r} ( 0 )}{\partial \omega} \right|_{\omega=0}^{} = 0$,
 and for the four-vertex $\widetilde{\Gamma}_{m_3 \sigma_3 ; m_4 \sigma_4 }^{m_1 \sigma_1 ; m_2 \sigma_2} ( 0, 0, 0, 0)$
(see Appendix \ref{secA:CT}).

We can evaluate the values of the quasiparticle parameters,
$\tilde{\epsilon}_{\rm d}^{}, \widetilde{\Gamma}, \widetilde{U}, \widetilde{W}$, and $\widetilde{J}$,
with use of the numerical renormalization group approach,
\cite{Hewson2004,PhysRevB.82.115123,PhysRevLett.108.056402}
because the quantities are defined at $T=0$ and $V=0$
(see Appendix \ref{secA:CRP}).
The nonequilibrium effect at low bias voltages $V \ll T_{\rm K}^{} \sim \widetilde{\Gamma}$
enters via perturbation expansion in the renormalized interactions.

The perturbation expansion up to only the second order in the renormalized interactions
gives a precise expression of the self-energy at $T=0$ up to the second order in $\omega$ and $V$,
and those of currents and current correlations up to order $V^3$.
The counter terms cancel
all the terms with higher orders in the renormalized interactions.
\cite{doi:10.1143/JPSJ.74.110,PhysRevB.83.075440,PhysRevB.83.241301,PhysRevLett.108.266401}
We shall calculate current cross-correlation
by the perturbation expansion in the renormalized interactions.

\subsection{Full counting statistics}

To calculate currents and current correlations,
we make use of the full counting statistics.
\cite{PROP:PROP200610305,RevModPhys.81.1665}
There are two major advantages of the full counting statistics.
One is that the technique gives all-order current correlations at once.
The other is that labeled counting fields
in the partition function classify scattering processes in the current,
which provides us an intuitive understanding of
the underlying physics of the currents and the current fluctuations.
Thus, the combination of the renormalized perturbation expansion
and the full counting statistics can be a powerful tool
to explore the mechanism that leads 
to entangled states in the current.

Let us start with the probability distribution $P( {\bm q})$
of the transferred charge
\begin{eqnarray}
{\bm q}=
\left(
q_{L1\uparrow }^{}, q_{L1\downarrow }^{}, q_{L2\uparrow }^{}, q_{L2\downarrow }^{},
q_{R1\uparrow }^{}, q_{R1\downarrow }^{}, q_{R2\uparrow }^{}, q_{R2\downarrow }^{}
\right)
\nonumber
\end{eqnarray}
with orbital $m$ spin $\sigma$
from lead $\alpha$ to the dot
in a time interval ${\cal T}$ between $t=-\frac{\cal T}{2}$ and $t=\frac{\cal T}{2}$,
which can provide the current correlation function to all orders.
In this paper, we consider the steady currents
in the long time limit ${\cal T} \to \infty$.
For simplification of the spin subscript,
$\sigma=\uparrow$ indicates spin angled to the $\theta (\phi)$ direction, and
$\sigma=\downarrow$ indicates spin angled to
the $\theta+\pi (\phi+\pi)$ direction
for orbital $m=1(2)$, 
in the following. 
This probability distribution is given by
\begin{eqnarray}
	P ( {\bm q})=
	\left\langle
	\prod_{m\sigma\alpha} \delta_{ \hat{n}_{\alpha m\sigma}^{}
	\left( - \frac{\cal T}{2}\right) - \hat{n}_{\alpha m\sigma}^{} \left( \frac{\cal T}{2}\right) }^{q_{\alpha m\sigma}^{}}
	\right\rangle \, ,
\end{eqnarray}
in terms of the operator for the number of charge in lead $\alpha$,
\begin{eqnarray}
\hat{n}_{\alpha m\sigma}^{} (t )
= \int_{-D}^{D} d\varepsilon \,
c_{\varepsilon \alpha m \sigma}^{\dagger} (t) \, c_{\varepsilon \alpha m \sigma}^{} (t)
\,.
\end{eqnarray}
The generation function for this probability distribution is given in the form
\begin{eqnarray}
 \chi \left( {\bm \lambda} \right) =  \sum_{\bm q} e^{i{\bm \lambda} \cdot {\bm q}} P\left( {\bm q}\right)
\end{eqnarray}
with the counting field
\begin{eqnarray}
	{\bm \lambda}=
	\left( 
	\lambda_{L1\uparrow}^{}, \lambda_{L1\downarrow}^{}, \lambda_{L2\uparrow}^{}, \lambda_{L2\downarrow}^{},
	\lambda_{R1\uparrow}^{}, \lambda_{R1\downarrow}^{}, \lambda_{R2\uparrow}^{}, \lambda_{R2\downarrow}^{}
	\right).
\nonumber 
\end{eqnarray}
The cumulant generating function can be written as
\begin{eqnarray}
\ln \chi ( {\bm \lambda} ) = \ln {\cal Z} ( {\bm \lambda} )
\, ,
\end{eqnarray}
in term of the partition function ${\cal Z} \left( {\bm \lambda} \right)$ for an extended Lagrangean
\begin{eqnarray}
{\cal L}_{\rm A}^{\mu} ( t, {\bm \lambda} ) =
{\cal L}_0^{\mu} ( t )
+ {\cal L}_{\rm T}^{\mu} ( t, {\bm \lambda} )
+ {\cal L}_{\rm I}^{\mu} ( t )
\, .
\label{eq:AIMwithCF}
\end{eqnarray}
This extended tunneling part is given by
\begin{eqnarray}
	&& {\cal L}_{\rm T}^{\mu} ( t , {\bm \lambda }) =
	\sum_{\alpha m \sigma} \left[
	v_{\alpha}^{} e^{ i\lambda_{\alpha m\sigma}^{\mu} } d_{m\sigma}^{\dagger} ( t) \psi_{\alpha m \sigma}^{} ( t)
	+ {\rm H.c.} \right]
	\, ,
	\label{eq:ExtnddTnnlngL}
\end{eqnarray}
where the sign of the counting field depends on the Keldysh contour as
$\lambda_{\alpha m\sigma}^{\mu} =
\left( \sigma_3^{} \right)_{}^{\mu\mu} \lambda_{m\sigma }^{}$.
Thus, the partition function is written in the path integral form:
\begin{eqnarray}
	&& {\cal Z} ( {\bm \lambda} ) \nonumber \\
	&& = \int 
	{\cal D}\left(\bar{c}_{\varepsilon \alpha  m\sigma}^{} \right) {\cal D} \left(c_{\varepsilon \alpha m\sigma}^{} \right)
	{\cal D}\left(\bar{d}_{m\sigma}^{} \right) {\cal D} \left(d_{m\sigma}^{} \right)
	e^{i{\cal S} ( {{\bm \lambda} })}
	\, ,
	\nonumber \\
\end{eqnarray}
where the action is given by
\begin{eqnarray}
{\cal S} ( {\bm \lambda }) &=&
\int_{C_{\cal T}^{}}^{} dt \, 
{\cal L}_{\rm A}^{\mu} ( t , {\bm \lambda })
\, ,
\end{eqnarray}
and $C_{\cal T}^{}$ is a Keldysh contour in a time interval ${\cal T}$
[see Fig. \ref{fig:KC}(b)].
The counting field is introduced
such that the terms where one electron moves from the left or right lead to the dot
gain one counting field
$e^{\lambda_{\alpha m\sigma}^{}}$,
as seen in the extended tunneling Lagrangean (\ref{eq:ExtnddTnnlngL}).
It is useful
for analyzing scattering processes in currents.

\section{Result and discussion}\label{sec:result}

Let us calculate the cumulant generating function for the current through the quantum dot in the Kondo regime.
Then, using the obtained cumulant generating function, current cross-correlations, averaged currents and shot noise, are derived
in terms of renormalized parameters, and we investigate spin entanglement of quasiparticle pairs emerging in the current. 
Finally, we discuss interaction dependence of these quantities with use of the numerical renormalization group approach.

\subsection{Cumulant generating function}
Applying the renormalized perturbation theory,
we can precisely include the local Fermi liquid properties in 
the partition function for low energies,
which can be written in the quasiparticle picture as
\begin{eqnarray}
	{\cal Z} ( {\bm \lambda})
	= \int
	{\cal D}\left(\tilde{\bar{d}}_{m\sigma}^{} \right) {\cal D} (\tilde{d}_{m\sigma}^{} )
	e^{i {\cal S}_{\rm red}^{} ( {{\bm \lambda} })}
	\, ,
	\label{eq:PrttnFnctn-FCSRPT}
\end{eqnarray}
where the reduced action is given by
\begin{eqnarray}
	&& {\cal S}_{\rm red}^{} ( {\bm \lambda} ) \nonumber \\
	&& =
	\int_{-\frac{\cal T}{2}}^{\frac{\cal T}{2}} dt \int_{-\frac{\cal T}{2}}^{\frac{\cal T}{2}} dt'
	\sum_{m\sigma} \tilde{\bar{\bm d}}_{m\sigma}^{} ( t )
	\left[ \widetilde{\bm g}_{{\rm d}m\sigma}^{\lambda} ( t - t' ) \right]^{-1}
	\tilde{\bm d}_{m\sigma}^{} ( t' )
	\nonumber \\
	&& \qquad + \sum_{\mu} \int_{-\frac{\cal T}{2}}^{\frac{\cal T}{2}} dt \,
	( \sigma_3^{} )_{}^{\mu\mu} \left[ \widetilde{\cal L}_{\rm I}^{\mu} ( t)+  \widetilde{\cal L}_{\rm CT}^{\mu} ( t) \right]
	\,.
	\label{eq:ActionwithCF}
\end{eqnarray}
Here, $\widetilde{\bm g}_{{\rm d}m\sigma}^{\lambda} ( t)$ is the Green's function of the free quasiparticle
with the counting fields (see Appendix \ref{secA:GF4FQ}).

Applying the perturbation expansion in the renormalized interactions
to the partition function given in Eq. (\ref{eq:PrttnFnctn-FCSRPT}),
the cumulant generating function up to the third order in
$V$
is precisely calculated
in terms of the renormalized parameters:
\begin{eqnarray}
	\ln \chi ( {\bm \lambda} )  &=& \ln {\chi}_0^{} ( {\bm \lambda} )
	+ \frac{\cal T}{2 \pi} V \left( \frac{V}{\widetilde{\Gamma}}\right)^2 \left( {\cal A} + {\cal B} \right) 
	+ {\cal O} ( V^5 )
	\, . \nonumber \\
	\label{eq:CGF-2ndodr}
\end{eqnarray}
This is a key result.
Equation (\ref{eq:CGF-2ndodr}) enables us
to calculate the current correlations in all orders
by differentiating it with respect to the counting fields.
It also describes the scattering processes of the low-energy excited states in the current.
The first term of Eq. (\ref{eq:CGF-2ndodr}) describes the free-quasiparticles contribution,
which is resulted from the zeroth order term
of the perturbation expansion in the renormalized interaction as
\begin{eqnarray}
	\ln \chi_0^{} ( {\bm \lambda} ) = \frac{{\cal T}}{2\pi}  \sum_{m\sigma} \int_{-\frac{V}{2}}^{+\frac{V}{2}} d \omega
	\ln \left[ 1+ T_{m\sigma} ( \omega ) \left( e^{ i \bar{\lambda}_{m\sigma}^{}} -1\right)\right]
	\, . \nonumber \\
\end{eqnarray}
In the free-quasiparticle's process, the quasiparticles are scattered by the resonant level near the Fermi level,
described by the transmission probability
\begin{eqnarray}
	T_{m \sigma} ( \omega ) = -\Gamma {\rm Im} G_{{\rm d}m\sigma}^{\rm r} (\omega )
	=\frac{\widetilde{\Gamma}^2}{\omega^2 + \widetilde{\Gamma}^2} \, .
\end{eqnarray}
The second term of Eq. (\ref{eq:CGF-2ndodr}) is obtained by second order calculation in the renormalized interactions as
\begin{widetext}
\begin{eqnarray}
	{\cal A}  &=& 
	\tilde{u}^2 \left\{
	\frac{1}{12} \sum_{m\sigma} \left( e^{-i \bar{\lambda}_{m\sigma}^{}}-1\right)
	+ \frac{1}{3} \sum_{m} \left[ e^{-i \left( \bar{\lambda}_{m\uparrow}^{} + \bar{\lambda}_{m\downarrow}^{} \right)}-1\right]
	\right\}
	\nonumber \\
	&& \qquad + 
	\frac{1}{4} \tilde{j}^2 \left[ \cos^2 \left( \theta - \varphi \right) + 1 \right]
	\left\{ \frac{1}{24} \sum_{m\sigma} \sum_{m'}^{m' \neq m}
	\left[ e^{-i \left(\lambda_{Rm\sigma }^{} -\lambda_{Rm' \sigma }^{}
	- \lambda_{Rm \bar{\sigma}}^{}
	+ \lambda_{Lm' \bar{\sigma}}^{} \right)}-1\right] \right. \nonumber \\
	&& \qquad \qquad + \frac{1}{24} \sum_{m\sigma} \sum_{m'}^{m' \neq m}
	\left[ e^{-i \left(\lambda_{L m\sigma }^{} +\lambda_{L m' \bar{\sigma} }^{}
	- \lambda_{L m \bar{\sigma}}^{}
	- \lambda_{R m' \sigma}^{} \right)}-1\right]
	+ \left. \frac{1}{3} \sum_{\sigma}
	\left[ e^{-i \left(\lambda_{L 2\sigma}^{}
	+ \lambda_{L 1\bar{\sigma}}^{}
	 - \lambda_{R 1\sigma}^{}
	 -\lambda_{R 2\bar{\sigma} }^{} \right)} -1\right] \right\} \nonumber \\
	&& \qquad + 
	\left[ \tilde{w} - \frac{1}{2} \tilde{j} \cos \left( \theta -\varphi\right) \right] ^2
	\left\{
	\frac{1}{6} \sum_{m\sigma} \left( e^{-i \bar{\lambda}_{m\sigma}^{}}-1\right)
	+ \frac{1}{3} \sum_{\sigma \sigma'} \left[ e^{-i \left( \bar{\lambda}_{1\sigma}^{} + \bar{\lambda}_{2\sigma'}^{} \right)}-1\right]
	\right\}
	\nonumber \\
	&& \qquad + 
	\tilde{j} \tilde{w} \cos\left( \theta - \varphi \right) 
	\left\{ \frac{1}{6} \sum_{m\sigma} \left( e^{-i \bar{\lambda}_{m\sigma}^{}} -1\right)
	+ \frac{2}{3} \sum_{\sigma} \left[ e^{-i \left( \bar{\lambda}_{1\sigma}^{} + \bar{\lambda}_{2\sigma}^{} \right)} -1\right]
	\right\}
	\, ,
	\\
	{\cal B} &=&
	\frac{1}{4} \tilde{j}^2 \left[ \cos \left( \theta - \varphi \right) - 1 \right]^2
	\sum_{\sigma}
	\left\{
	\frac{1}{3} \left[ e^{-i \left( \lambda_{L 1\sigma}^{} + \lambda_{L 2\sigma}^{} - \lambda_{R 1\bar{\sigma}}^{} -\lambda_{R 2\bar{\sigma}}^{} \right) } -1 \right]
	+\frac{1}{24} \left[e^{-i \left(\lambda_{R 1\sigma}^{} + \lambda_{L 2\sigma}^{} - \lambda_{R 1\bar{\sigma}}^{} -\lambda_{R 2\bar{\sigma}}^{} \right) } -1 \right]
	\right. \nonumber \\
	&& \qquad  \qquad +
	\frac{1}{24} \left[ e^{-i \left(\lambda_{L 1\sigma}^{} + \lambda_{R 2\sigma}^{} - \lambda_{R 1\bar{\sigma}}^{} -\lambda_{R 2\bar{\sigma}}^{} \right) } -1 \right]
	+\frac{1}{24} \left[ e^{-i \left(\lambda_{L 1\sigma}^{} + \lambda_{L 2\sigma}^{} - \lambda_{R 1\bar{\sigma}}^{} -\lambda_{L 2\bar{\sigma}}^{} \right) } -1 \right]
	\nonumber \\
	&& \qquad  \qquad \left.
	+ \frac{1}{24} \left[ e^{-i \left(\lambda_{L 1\sigma}^{} + \lambda_{L 2\sigma}^{} - \lambda_{L 1\bar{\sigma}}^{} -\lambda_{R 2\bar{\sigma}}^{} \right) } -1 \right]
	\right\} \nonumber \\
	&& \qquad +
	 \frac{1}{4} \tilde{j}^2 \sin^2 \left( \theta - \varphi \right) \sum_{mm'\sigma'} \sum_{m'}^{m' \neq m}
	\left\{
	\frac{1}{3}
	\left[ e^{i \left( \lambda_{R m {\sigma}' }^{} + \lambda_{R m' {\sigma}}^{} -\lambda_{L m' \bar{\sigma}}^{} - \lambda_{L m {\sigma}'}^{}\right)} -1 \right]
	+ \frac{1}{12}
	\left[ e^{i \left( \lambda_{R m {\sigma}}^{} - \lambda_{L m \bar{\sigma}}^{} \right)} -1 \right]
	\right. \nonumber \\
	&& \qquad \qquad \qquad \qquad \left.
	+ \frac{1}{24}
	\left[ e^{i \left( \lambda_{R m {\sigma}' }^{} + \lambda_{R m' {\sigma}}^{} -\lambda_{R m' \bar{\sigma}}^{} - \lambda_{L m {\sigma}'}^{}\right)} -1 \right]
	+ \frac{1}{24}
	\left[ e^{i \left( \lambda_{R m {\sigma}' }^{} + \lambda_{L m' {\sigma}}^{} -\lambda_{L m' \bar{\sigma}}^{} - \lambda_{L m {\sigma}}^{}\right)} -1 \right]
\right\}
\, ,
\end{eqnarray}
\end{widetext}
with
$\bar{\sigma} =\uparrow \left( \downarrow \right)$
for $\sigma =\downarrow \left( \uparrow \right)$
and $\bar{\lambda}_{m \sigma}^{} = \lambda_{L m \sigma}^{} - \lambda_{R m \sigma}^{}$.
Here the renormalized interactions scaled by the renormalized linewidth,
$\tilde{u} := \frac{\widetilde{U}}{\pi\widetilde{\Gamma}},
\tilde{w} := \frac{\widetilde{W}}{\pi\widetilde{\Gamma}}$.
and
$\tilde{j} := \frac{\widetilde{J}}{\pi\widetilde{\Gamma}}$,
express the strengths of the interactions of the Fermi liquid.
${\cal A}$ and ${\cal B}$ represent the contribution of the interacting quasiparticles,
and it is peculiar to nonequilibrium beyond the linear response current.

At the end of this subsection,
we mention properties of the free-quasiparticle term. 
For low bias voltages, 
this term can be expanded in bias voltage $V$ up to the third order as
\begin{eqnarray}
	&& \ln \chi_0^{} ( {\bm  \lambda} ) \nonumber  \\
	&& \ \sim \frac{\cal T}{2\pi} V \sum_{m\sigma}
	\left[
	i \bar{\lambda}_{m\sigma}^{} 
	 - \frac{1}{12}
	\left( \frac{V}{\widetilde{\Gamma}} \right)^2
	\left( e^{ - i  \bar{\lambda}_{m\sigma}^{} } - 1 \right) \right] \nonumber \\
	&& \qquad + {\cal O} ( V^5) \, .
\end{eqnarray}
This term
does not contain any current correlations between the two orbitals.
It is natural because this term is accounted for by
the free quasiparticles.
Therefore, only quasiparticles excited by the renormalized interactions in the nonlinear current of order $V^3$
can form spin-entanglement between  the two channels.

We note that there is no current of order $V^2$
because of our setting of the particle-hole symmetry.

\subsection{Current and current-correlations}

\paragraph{Cross-correlation}
We first calculate interorbital cross-correlations of current fluctuations,
which can be readily derived as a derivative of Eq. (\ref{eq:CGF-2ndodr})
with respect to counting fields:
\begin{eqnarray}
	C_{\alpha\alpha}^{} ( \theta, \varphi)
	&=& \int_{- \infty}^{\infty} dt \ \left\langle
	\delta I_{\alpha1\theta}^{} ( t) \, \delta I_{\alpha 2\varphi}( 0)
	\right\rangle
	\nonumber \\
	&=& \frac{e^2}{\cal T} \left( -i \right)^2 \left. 
	\frac{\partial}{\partial \lambda_{\alpha 1\uparrow }^{}} \frac{\partial}{\partial \lambda_{\alpha 2\uparrow }^{}}
	 \ln \chi ( {\bm \lambda})
	\right|_{{\bm \lambda} = 0}^{} 
	\nonumber \\
	&=& B_1^{} - B_2^{} \cos ( \theta - \varphi ) + {\cal O} \left( V^5 \right)
	\label{eq:cccorrelation4twstangl}
\end{eqnarray}
with
\begin{eqnarray}
	B_{1}^{} &=&
	\frac{1}{2\pi} V\left( \frac{V}{\widetilde{\Gamma}} \right)^2
	\left( \frac{1}{4} \tilde{j}^2
	+ \frac{1}{3} \tilde{w}^2 \right)
	\, , \\
	B_{2}^{} &=& 
	\frac{1}{2\pi} V\left( \frac{V}{\widetilde{\Gamma}} \right)^2
	\left(
	\frac{1}{4} \tilde{j}^2 - \frac{1}{3} \tilde{w}  \tilde{j}
	\right) \, .
\end{eqnarray}
Here, the operator of current fluctuation 
can be defined by
$\delta I_{\alpha m \sigma}^{}:= I_{\alpha m \sigma}^{} -\left\langle I_{\alpha m \sigma}^{} \right\rangle$.
This is one of the main result of this paper.
Here, the angle independent and dependent
terms of Eq. (\ref{eq:cccorrelation4twstangl})
are related to the charge and spin correlation of excited particles 
and holes in the current, as following.
The cross-correlation of charge currents and spin currents 
between orbitals $m=1$ and $2$ are given by
\begin{eqnarray}
	C_{\alpha\alpha}^{\rm c}
	&=& \int_{- \infty}^{\infty} dt \ \left\langle
		\delta I_{\alpha1}^{\rm c} ( t) \delta I_{\alpha 2}^{\rm c}( 0)
		\right\rangle
		\nonumber \\
	&=& C_{\alpha\alpha}^{} ( \theta, \varphi ) + C_{\alpha\alpha}^{} ( \theta+ \pi, \varphi )
	\nonumber \\
	&& \quad + C_{\alpha\alpha}^{} ( \theta, \varphi + \pi ) + C_{\alpha\alpha}^{} ( \theta
	+ \pi, \varphi + \pi )
	\nonumber \\
	&=& 4 B_1^{} \, , \\
	C_{\alpha\alpha}^{\rm s}( \theta, \varphi)
	&=& \int_{- \infty}^{\infty} dt \ \left\langle
		\delta I_{\alpha1\theta}^{\rm s} ( t )
		\delta I_{\alpha 2\varphi}^{\rm s} ( 0 )
		\right\rangle \nonumber \\
	&=& C_{\alpha\alpha}^{} ( \theta, \varphi )
	- C_{\alpha\alpha}^{} ( \theta+\pi, \varphi )
	\nonumber \\
	&& \quad - C_{\alpha\alpha}^{} ( \theta, \varphi+\pi )
	+ C_{\alpha\alpha}^{} ( \theta+\pi, \varphi+\pi )
	\nonumber \\
	&=& 4 B_2^{} \cos ( \theta - \varphi ) \, ,
\end{eqnarray}
respectively. Here,
\begin{eqnarray}
	I_{\alpha m}^{\rm c} &=& \sum_{\sigma} I_{\alpha m \sigma}^{} 
	\, , \\
	I_{\alpha m \theta}^{\rm s} &=& I_{\alpha m \theta}^{} - I_{\alpha m \, \theta+\pi}^{}
	\, ,
\end{eqnarray}
are the charge and spin current, respectively.
In our model,
every single quasiparticle and hole carries both a charge and a spin.
Thus, the ratio,
\begin{eqnarray}
	\frac{C_{\alpha\alpha}^{\rm s} \left( \theta, \varphi \right)}{ C_{\alpha\alpha}^{\rm c}}
	= \frac{B_2^{}}{B_{1}} \cos \left( \theta - \varphi\right) \, ,
	\label{eq:cc-effctv-spn}
\end{eqnarray}
corresponds to the cross-correlation for effective spins per current-carrying charge.
The prefactor
is determined by the local-Fermi-liquid parameters, 
specifically the interorbital residual interactions $\tilde{w}$ and $\tilde{j}$ as,
\begin{eqnarray}
	{\cal R} :=
	\frac{B_2^{}}{B_1^{}}
	=\frac{1 - \frac{4}{3} \left( \frac{\tilde{w}}{\tilde{j}}\right)}
	{1 + \frac{4}{3} \left( \frac{\tilde{w}}{\tilde{j}}\right)^2}
	\, .
\end{eqnarray}
Since ${\cal R}$ is the value of Eq. (\ref{eq:cc-effctv-spn}) for $\theta = \varphi$,
positive (negative) values of ${\cal R}$ indicate
that the charge pairs with the parallel (antiparallel) spins are dominant in the current.

We also find that the
cross-correlation of the current with three different angular momenta
is induced by the exchange interaction:
\begin{eqnarray}
	C_{\rm three}^{}
	&=& \int dt \int dt' \left\langle
	\delta I_{\alpha m\sigma }^{} ( t)
	\delta I_{\alpha m'\sigma }^{} ( t')
	\delta I_{\alpha m'\bar{\sigma} }^{} ( 0) \right\rangle
	\nonumber \\
	&=& \frac{( - i)^3 }{\cal T}
	\left.
	\frac{\partial}{\partial\lambda_{\alpha m\sigma }^{}}
	\frac{\partial}{\partial\lambda_{\alpha m'\sigma }^{}}
	\frac{\partial}{\partial\lambda_{\alpha m' \bar{\sigma} }^{}}
	\ln \chi ( {\bm \lambda } )
	\right|_{{\bm \lambda}=0}^{}
	\nonumber \\
	&=& \frac{1}{2\pi} \frac{1}{24}  V\left( \frac{V}{\widetilde{\Gamma}} \right)^2 \tilde{j}^2 \, .
	\label{eq:3CrrntCC}
\end{eqnarray}
There is a cross-correlation of four different channels.
However,  it is equivalent to the above cross-correlation
of three spin-orbit channels 
because of 
the  conservation of the total angular momentum 
consisting of the spin and orbital angular momenta
during the quasiparticle scattering processes  by the residual interactions.

We shall
demonstrate the existence of
the spin-entanglement between the orbitals.
The Lagrangean for quasiparticle's interaction given by Eq.\,(\ref{eq:lagrangean4qpInt})
can be rewritten in terms of the spin-singlet and triplet components:
\begin{eqnarray}
	\widetilde{\cal L}_{\rm I}^{\mu} (t)
	&=& \widetilde{U} \sum_m  \tilde{n}_{{\rm d}m\uparrow}^{\mu} ( t ) \,
	\tilde{n}_{{\rm d}m\downarrow}^{\mu} ( t )
	+ \widetilde{W} \tilde{n}_{{\rm d}m}^{\mu} ( t ) \,
	\tilde{n}_{{\rm d}m}^{\mu} ( t )
	\nonumber \\ 
	&& \quad 
	-\frac{3\widetilde{J}}{2} \widetilde{J} \, \bar{b}_{\rm s}^{\mu} ( t ) \, b_{\rm s}^{\mu} ( t )
	+ \frac{\widetilde{J}}{2}  \sum_{i=0,\pm}^{} \bar{b}_{{\rm t}i}^{\mu} ( t ) \, b_{{\rm t}i}^{\mu} ( t)
	\, ,
	\label{eq:exctdstt}
\end{eqnarray}
where
\begin{eqnarray}
	\bar{b}_{\rm s}^{\mu} ( t)
	&=& - \frac{1}{\sqrt{2}} \left[
	\tilde{\bar{d}}_{1\uparrow}^{\mu} ( t) \tilde{\bar{d}}_{2\downarrow}^{\mu} ( t)
	- \tilde{\bar{d}}_{1\downarrow}^{\mu} ( t) \tilde{\bar{d}}_{2\uparrow}^{\mu} ( t) \right] 
	\, , \nonumber \\
	b_{\rm s}^{\mu} ( t)
	&=& \frac{1}{\sqrt{2}} \left[
	\tilde{d}_{1\uparrow}^{\mu} ( t) \tilde{d}_{2\downarrow}^{\mu} ( t)
	- \tilde{d}_{1\downarrow}^{\mu} ( t) \tilde{d}_{2\uparrow}^{\mu} ( t) \right] 
	\, , \label{eq:cdgn4snglt} \\
	\bar{b}_{{\rm t}0}^{\mu} ( t)
	&=& - \frac{1}{\sqrt{2}} \left[
	\tilde{\bar{d}}_{1\uparrow}^{\mu} ( t) \tilde{\bar{d}}_{2\downarrow}^{\mu} ( t)
	+ \tilde{\bar{d}}_{1\downarrow}^{\mu} ( t) \tilde{\bar{d}}_{2\uparrow}^{\mu} ( t) \right] 
	\, , \nonumber \\
	b_{{\rm t}0}^{\mu} ( t)
	&=& \frac{1}{\sqrt{2}} \left[
	\tilde{d}_{1\uparrow}^{\mu} ( t) \tilde{d}_{2\downarrow}^{\mu} ( t)
	+ \tilde{d}_{1\downarrow}^{\mu} ( t) \tilde{d}_{2\uparrow}^{\mu}( t) \right]
	 \, , \label{eq:cdgn4trplt0} \\
	\bar{b}_{{\rm t}+}^{\mu} ( t)
	&=& - \tilde{\bar{d}}_{1\uparrow}^{\mu} ( t) \tilde{\bar{d}}_{2\uparrow}^{\mu} ( t)
	\, , \ 
	b_{{\rm t}+}^{\mu} ( t)
	= \tilde{d}_{1\uparrow}^{\mu} ( t) \tilde{d}_{2\uparrow}^{\mu} ( t) \, ,\\
	\bar{b}_{{\rm t}-}^{\mu} ( t)
	&=& - \tilde{\bar{d}}_{1\downarrow}^{\mu} ( t) \tilde{\bar{d}}_{2\downarrow}^{\mu} ( t)
	\, , \ 
	b_{{\rm t}-}^{\mu} ( t)
	= \tilde{d}_{1\downarrow}^{\mu} ( t) \tilde{d}_{2\downarrow}^{\mu} ( t) \, ,
	\label{eq:cdgn4trplt1} 
\end{eqnarray}
are the Grassmann number for the spin-singlet and triplet states of two generated particles (holes)  between the orbitals.
Applying the perturbation expansion with respect to $\widetilde{\cal L}_{\rm I}^{\mu} (t)$,
the term of the generating function with
order $\widetilde{J}^2$
can be written by Green's function for the singlet state and the triplet states:
\begin{widetext}
	\begin{eqnarray}
		\frac{\widetilde{J}^2}{4} \sum_{\mu\nu} (\sigma_3^{})^{\mu\mu} (\sigma_3^{})^{\nu\nu}
		\int_{- \frac{\cal T}{2}}^{+\frac{\cal T}{2}} dt \, dt'
		\left[
		9\left\langle
		b_{\rm s}^{\mu} (t) \bar{b}_{\rm s}^{\nu} (t') 
		\right\rangle
		\left\langle
		\bar{b}_{\rm s}^{\mu} (t) b_{\rm s}^{\nu} (t') 
		\right\rangle
		+ \sum_{i=0,1}
		\left\langle
		b_{{\rm t}i}^{\mu} (t) \bar{b}_{{\rm t}i}^{\nu} (t') 
		\right\rangle
		\left\langle
		\bar{b}_{{\rm t}i}^{\mu} (t) b_{{\rm t}i}^{\nu} (t') 
		\right\rangle
		\right] \, ,
		\label{eq:qp4intb}
	\end{eqnarray}
\end{widetext}
where
\begin{eqnarray}
	\left\langle \cdots \right\rangle
	=\frac{\int {\cal D}\left(\tilde{\bar{d}}_{m\sigma}^{} \right) {\cal D} (\tilde{d}_{m\sigma}^{} )  \cdots e^{i{\cal S}_{\rm red}^{(0)}}}{\int {\cal D}\left(\tilde{\bar{d}}_{m\sigma}^{} \right) {\cal D} (\tilde{d}_{m\sigma}^{} ) e^{i{\cal S}_{\rm red}^{(0)}}}
\end{eqnarray}
is an expectation value.
Equation (\ref{eq:qp4intb}) shows that the spin entangled pairs
given by Eqs. (\ref{eq:cdgn4snglt})-(\ref{eq:cdgn4trplt1}) are excited
and result in the cross-correlation between the two currents.
We note that there are additional $\widetilde{J}^2$ terms due to the counter term.
 However, they only eliminate the overcounting of the renormalization effect,
 and do not affect the form of the entangled pairs.

\paragraph{Averaged current}
The average current in each channel of the right lead is also calculated
from the first derivative of Eq. (\ref{eq:CGF-2ndodr}):
\begin{eqnarray}
	\left\langle I_{\alpha m \sigma}^{} \right\rangle
	&=& \left. \frac{e}{\cal T} \left( -i \right)\frac{\partial}{\partial \lambda_{\alpha m\sigma }^{}}
	\ln \chi  \left({\bm \lambda} \right) \right|_{{\bm \lambda}=0}^{} \nonumber \\
	& =& \left(1-2\delta_{\alpha }^{R}\right)
	\left( 
	I_u^{} -  \left\langle I_{\alpha m\sigma}^{\rm NL} \right\rangle
	\right) \, ,
\label{eq:current}
\end{eqnarray}
where
$I_u^{} = \frac{1}{2\pi}V$
is the linear current, and
\begin{eqnarray}
	 \left\langle I_{\alpha m\sigma}^{\rm NL} \right\rangle
	= \frac{1}{2\pi} \frac{1+ 5\widetilde{\cal I}}{12} V \left( \frac{V}{\widetilde{\Gamma}} \right)^2
	+ {\cal O} \left( V^5 \right)
\end{eqnarray}
is the nonlinear current with
$\widetilde{\cal I}=\tilde{u}^2+2\tilde{w}^2+\frac{3}{2}\tilde{j}^2$.
The averaged current is independent of the observation angles $\theta$ and $\varphi$.

\paragraph{Shot noise}
The shot noise is a current noise due to the charge discretization, 
and  is simply given by 
 auto-correlation of the full current 
 through the Anderson impurity.
This is 
because the noise source
is only a small amount of the scattering state of the excited quasiparticles
by the renormalized interaction in the current.
Thus, the shot noise is given by
the second derivation with respect to the counting field as
\begin{eqnarray}
	S &:=& \int dt \, \left\langle
	\delta I \left( t\right) \delta I \left( 0\right) + \delta I \left( 0\right) \delta I \left( t\right)
	\right\rangle
	\nonumber \\
	&=& \left( -i \right)^2 \frac{2 e^2}{\cal T}
	\left.
	\frac{\partial^2}{\partial \lambda^2}
	\ln \chi \left( \lambda \right)
	\right|_{\lambda=0}
	\nonumber \\
	&=& \frac{1}{3\pi} V\left( \frac{V}{\widetilde{\Gamma}} \right)^2 \left( 1 + 9 \widetilde{\cal I} \right)
	\, ,
\end{eqnarray}
where
\begin{eqnarray}
	\chi \left( \lambda \right)
	= \left. \chi \left( {\bm \lambda} \right) \right|_{\lambda_{Lm\sigma}^{} = -\lambda_{Rm\sigma}^{}=\frac{\lambda}{2}}^{}
	\, .
\end{eqnarray}
This form of the shot noise has 
already been derived for Hund's rule exchange interaction $J<0$,
\cite{PhysRevLett.108.266401}
and here it is naturally extended  to the whole of the
local-Fermi-liquid region, including antiferromagnetic interaction $J>0$.
The Fano factor for the backscattering current is given
by the ratio of the shot noise and the nonlinear current as
\begin{eqnarray}
	F_{\rm b}^{}
	:=\frac{S}{2e\left\langle I_{\alpha m\sigma }^{\rm NL}\right\rangle}
	=\frac{1+9\widetilde{\cal I}}{1 + 5 \widetilde{\cal I}}.
\end{eqnarray} 
The ratio of the current and its own shot noise usually gives an effective 
charge of the current-carrying state.
However, a variety of quasiparticle-hole pairs are excited in the current  
owing to the continuous low-energy spectrum
of the local Fermi-liquid,
 as seen in Eq. (\ref{eq:CGF-2ndodr}).
In this case,
 the Fano factor can be defined as
 an average of the current-carrying effective charges $e_i^* (i=1,2, \cdots)$
 of form
 \cite{PhysRevLett.97.086601}
 \begin{eqnarray}
 	eF_{\rm b}^{}= \frac{\sum_{i}  \left\langle \left( e_i^* \right)^2 \right\rangle}
	{\sum_{i} \left\langle e_i^* \right\rangle}.
	\label{eq:MeaningofEffctvChrg}
 \end{eqnarray}

In the remainder of this section,
we investigate the interaction dependence of the transport quantities derived above,
using
the renormalized interactions calculated
with the numerical renormalization group.
In particular, we shall discuss
behaviors of the transport quantities for {\it ferromagnetic} ($J<0$)
 and {\it antiferromagnetic} ($J>0$) exchange interactions.

\subsection{Ferromagnetic exchange interaction}

\begin{figure}[tb]
	\includegraphics[width=7.0cm]{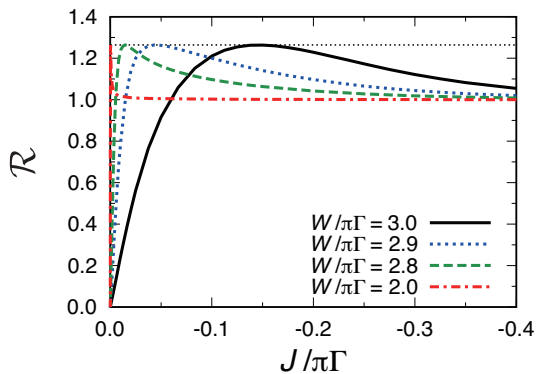}
	\caption{\label{fig:BB-ferro}
	The value of ${\cal R}$
	as a function of ferromagnetic $J (<0)$,
	with $U=3.0 \pi \Gamma$,
	and several choices of
	$W=3.0 \pi \Gamma,\, 2.9\pi \Gamma,\, 2.8\pi \Gamma$, and $2.0\pi \Gamma$.
	The thin dotted line indicates
	the maximum value ${\cal R} =\frac{1}{2}+ \frac{\sqrt{21}}{6}=1.263 \cdots$.
	}
\end{figure}

\begin{figure}[b]
	\includegraphics{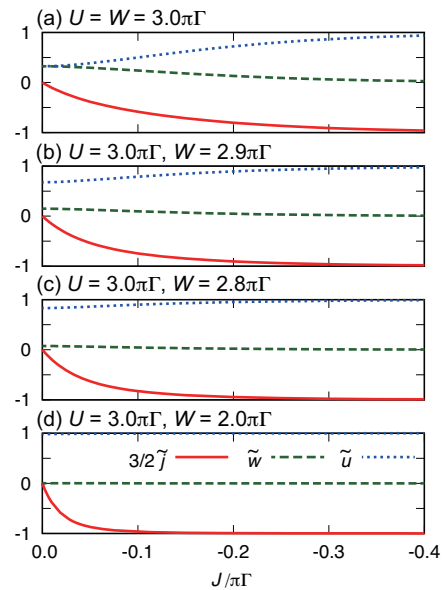}
	\caption{\label{fig:rnp-ferro}
	The renormalized interactions,
	$\tilde{j}$ (red solid line),
	$\tilde{u}$ (green dashed line),
	and $\tilde{w}$ (blue dotted line),
	as functions of ferromagnetic $J (<0)$,
	with $U=3.0 \pi \Gamma$,
	and several choices of
	(a) $W=3.0\pi \Gamma$, (b) $W=2.9\pi \Gamma$, (c) $W=2.8\pi \Gamma$, and (d) $W=2.0$.
	}
\end{figure}

Figure \ref{fig:BB-ferro} shows
${\cal R}$ as a function of ferromagnetic $J$
for four values of $W$ with $U$ fixed as $3.0\pi\Gamma$.
For $U=3.0\pi\Gamma$, the ground state
stays in the Kondo regime, and  
it evolves continuously as $W$ and $J$ vary.
The calculated renormalized interactions of
the local Fermi liquid, $\tilde{u}, \tilde{w}$, and $\tilde{j}$
are plotted in Fig. \ref{fig:rnp-ferro}
as  functions of
bare ferromagnetic exchange interaction $J$.

${\cal R}$ takes positive values for any finite strength of $J$,
which indicates that the charge pairs with parallel spins are always dominant in the current.
In the absence of the exchange interaction $J=0$,
the renormalized one is also $\tilde{j}=0$ 
and thus no spin angle-dependent interactions are induced,
while angle-independent renormalized interactions
 $\tilde{u}$ and $\tilde{w}$ are finite.
As a result, $B_2^{}=0$ while $B_1^{}$ is finite, 
and the ratio becomes ${\cal R}=0$ at $J=0$.
With increase of the strength of $J$, 
the spin entanglement between the two orbitals is enhanced.
Then, ${\cal R}$ takes a common maximum
$\frac{1}{2} + \frac{\sqrt{21}}{6}$ at a finite $J=J_{}^*$
for any finite strength of $W$ with $U>W$.
The two interorbital renormalized interactions,
$\tilde{j}$ and $\tilde{w}$,  
cooperatively enhance the spin entanglement
as the cross-term $\tilde{w}\tilde{j}$ arises in $B_2^{}$.
For larger ferromagnetic interaction $-J \gg \widetilde{\Gamma} \sim T_{\rm K}^{}$
and $U>W$,
the system crosses over to the $S=1$ full screening Kondo regime
where the renormalized interactions take the universal values of the $S=1$ Kondo fixed point:
$\tilde{u} \to 1, \tilde{w} \to 0$ ,
and $\tilde{j}\to-\frac{2}{3}$,
as seen in Fig. \ref{fig:rnp-ferro}.
\cite{PhysRevB.82.115123}
In this limit, charge fluctuations due to the interorbital Coulomb interaction
 are suppressed, i.e., $\tilde{w} \to 0$, 
and the spin fluctuations are maximally enhanced,
which results in
\begin{eqnarray}
	{\cal R} \to 1 \, .
\end{eqnarray}
We note that ${\cal R}=1$ does not mean that the current is fully spin-polarized.
As seen in the generating function given in Eq. (\ref{eq:CGF-2ndodr}), correlated quasiparticles and holes excited by the exchange interaction give rise to pure spin current and charge current.
Therefore, ${\cal R}=1$ simply means that the number of current-carrying spins is equal to that of charges.

The crossover 
from the SU(2) Kondo state for $U>W$ or the SU(4) Kondo state for $U=W$, and $J=0$
to the fully screened $S=1$ Kondo state
at large strength of $J$,
is observed in the ratio ${\cal R}$.
\cite{PhysRevLett.108.266401}
Note that,
for $W=0$,
the ratio is constant ${\cal R} = 1$ for nonzero $J$,
because the residual interorbital Coulomb interaction is always zero $\tilde{w}=0$,
$\widetilde{W}=0$ in this case.
As $-J$ increases,
the renormalized interactions and ${\cal R}$ more rapidly converge to 
their own universal values for smaller values of $W$,
because  the Kondo temperature decreases with  $W$.

This crossover through variation in the renormalized parameters
can also be seen in the cross-correlation of three currents given by (\ref{eq:3CrrntCC})
and shot noise.
The cross-correlation of three currents takes the form
\begin{eqnarray}
	C_{\rm three}^{} = \frac{1}{2\pi} \frac{1}{864} V \left( \frac{\pi V}{T_{\rm K}^{}} \right)^2
\end{eqnarray} 
in the $S=1$ Kondo limit ($J/\Gamma \to -\infty$).
The shot noise is also enhanced with the increase of two-particle backscattering
due to the renormalized exchange interaction.
This has been clearly seen in enhancement of the Fano factor.
\cite{PhysRevLett.108.266401}

\subsection{Antiferromagnetic interaction}
\begin{figure}[tb]
	\includegraphics[width=7.0cm]{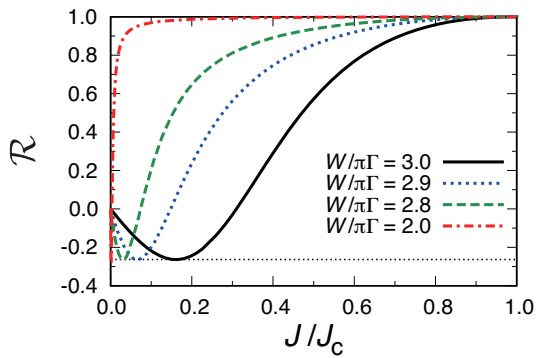}
	\caption{\label{fig:BB-antiferro}
	The values of ${\cal R}$
	as a function of antiferromagnetic
	$J/J_{\rm c}^{}$,
	with $U=3.0 \pi \Gamma$,
	and several choices of
	(a) $W=3.0\pi \Gamma,\, 2.9\pi \Gamma,\, 2.8\pi \Gamma$, and $2.0\pi \Gamma$.
	$J$ is the normalized by the critical value $J_{\rm c}^{}$ which is listed in Table \ref{tab:crtclJ}.
	The thin dotted line indicates
	the minimum value ${\cal R} = \frac{1}{2}- \frac{\sqrt{21}}{6}=-0.263 \cdots$.
	}
\end{figure}

\begin{figure}[tb]
	\includegraphics{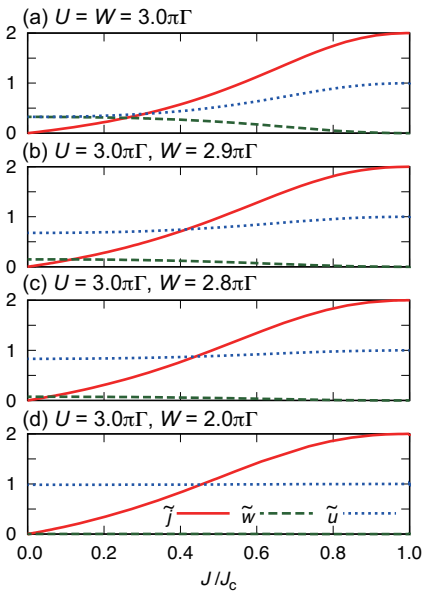}
	\caption{\label{fig:rnp-antiferro}
	The renormalized interactions,
	$\tilde{j}$ (red solid line)
	$\tilde{u}$ (green dashed line)
	and $\tilde{w}$ (blue dotted line)
	as functions of antiferromagnetic $J/J_{\rm c}^{}$,
	with $U=3.0 \pi \Gamma$,
	and several choices of
	(a) $W=3.0\pi \Gamma$, (b) $W=2.9\pi \Gamma$, (c) $W=2.8\pi \Gamma$, and (d) $W=2.0\pi \Gamma$.
	The exchange interaction is normalized by the critical value $J_{\rm c}^{}$ which is listed in Table \ref{tab:crtclJ}.
	}
\end{figure}

Figures \ref{fig:BB-antiferro} and \ref{fig:rnp-antiferro} show
the computed values of ${\cal R}$,
and the renormalized interactions, $\tilde{u},\, \tilde{w}$ and $\tilde{j}$,
respectively,
as a functions of ferromagnetic $J>0$
with $U=3\pi \Gamma$
and several values of $W$. 
There is a critical point at $J=J_{\rm c}^{}$
for antiferromagnetic $J(>0)$.
The values of the critical interaction $J_{\rm c}^{}$ are listed
in Table \ref{tab:crtclJ}.
\begin{table}[b]
	\caption{\label{tab:crtclJ}%
	The critical values $J_{\rm c}^{}$ calculated by the numerical renormalization group,
	for $U= 3.0\pi \Gamma$,
	and the values of $W$ used in Figs. \ref{fig:BB-antiferro} and \ref{fig:rnp-antiferro}.}
	\begin{ruledtabular}
	\begin{tabular}{ccccc}
	$\frac{W}{\pi\Gamma}$ &
	3.0 & 2.9 & 2.8 & 2.0\\
	$\frac{J_c^{}}{\pi\Gamma}$ &
	0.1140 & 0.0950 & 0.08121 & 0.04034
	\end{tabular}
	\end{ruledtabular}
\end{table}
For large antiferromagnetic interactions $J> J_{\rm c}^{}$,
two electrons occupied in the two orbitals form
an isolated singlet state and decouple
from the conduction electrons in the leads,
and  then  no electric currents  can flow through the quantum dot.
\cite{PhysRevLett.108.056402,PhysRevB.86.125134}
Therefore,
we focus on the region $J<J_{\rm c}^{}$
where the low-energy state is accounted for
by the local Fermi-liquid and electric current through the dot arises.

As the antiferromagnetic interaction increases,
${\cal R}$ decreases to a common and negative minimum
$\frac{1}{2} - \frac{\sqrt{21}}{6}$,
where cooperation of the interorbital interactions, $\tilde{w}$ and $\tilde{j}$,
maximizes the number  of the charge pairs with antiparallel spins in the current.
Then, ${\cal R}$ turns to increases
to the value ${\cal R} = 1$
at the limit $J \to J_{\rm c}^{} + 0_{}^-$ where
the renormalized interactions take the values of
$\tilde{u} \to 1, \tilde{w} \to 0$,
and $\tilde{j} \to 2$
for $U>W$
\cite{PhysRevLett.108.056402,PhysRevB.86.125134}
as seen in Fig. \ref{fig:rnp-antiferro}.
The number of the charge pairs with antiparallel spins decreases,
and ones with parallel spins become dominant in the current.
The excited state is described by Eq. (\ref{eq:exctdstt})
as long as the system stays in the local-Fermi-liquid region.
Note that,
for $W=0$,
the ratio is unity for nonzero $J$,
because the renormalized interorbital Coulomb interaction is always zero,
$\tilde{w}=0$.

\begin{figure}[tb]
	\includegraphics[width=7.0cm]{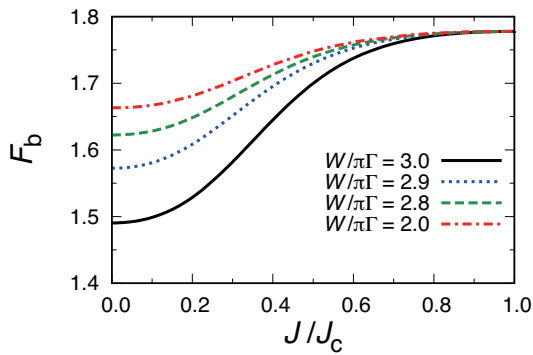}
	\caption{\label{fig:FF-antiferro}
	The Fano factor $F_b^{}$,
	as a function of antiferromagnetic
	$J/J_{\rm c}^{} (J>0)$,
	with $U=3.0 \pi \Gamma$,
	and several choices of
	$W=3.0\pi \Gamma, \, 2.9\pi \Gamma, \, 2.8\pi \Gamma$, and $2.0\pi \Gamma$.
	}
\end{figure}

The crossover can also be seen in the cross-correlation of three currents given by (\ref{eq:3CrrntCC}) and the shot noise
through variation in the renormalized parameters.
The cross-correlation of three currents in the limit $J \to J_{\rm c}^{}+0^-$ is given in a form
\begin{eqnarray}
	C_{\rm three}^{} = \frac{1}{2\pi}\frac{1}{96} V \left( \frac{\pi V}{T_{\rm K}^{}} \right)^2.
\end{eqnarray} 
The Fano factor of the shot noise of the nonlinear current
is shown in Fig. \ref{fig:FF-antiferro}
as a function of $J$.
At the limit $J \to J_{\rm c}^{} + 0_{}^-$,
the Fano factors take a universal value
\begin{eqnarray}
	F_{\rm b}^{} \to \frac{16}{9}.
\end{eqnarray}
By classifying the backscattered current by the effective charges,
\cite{PhysRevLett.108.266401}
the ratio of probability to generate current with effective charge $e$ and $2e$ are found to be 4:7,
which result in the value through Eq. (\ref{eq:MeaningofEffctvChrg}).
Even near the critical point $J \to J_{\rm c}^{} + 0_{}^-$,
the Kondo effect or the Fermi liquid state with strong renormalization are held as the low energy state.
Thus,
$2e$ charge states are strongly backscattered in the current,
which can be observed as an increase of $F_{\rm b}^{}$.
With an increase of $J$, the renormalized interactions, ${\cal R}$ and $F_{\rm b}^{}$, more rapidly converge
to their own universal values 
for smaller values of $W$,
because the Kondo temperature decreases with $W$, 
similarly to the case with the ferromagnetic interaction.

\section{Summary}\label{sec:summary}

We have investigated spin-entanglement of quasiparticle pairs 
excited by the renormalized interactions of the local Fermi liquid,
which arise in nonlinear currents
through the Kondo state of quantum dots with two degenerate orbitals and exchange interactions.
We have shown, the cumulant generating function of the current is precisely
described in the terms of 
the local Fermi-liquid parameters, up to third order in the applied bias voltage.
Using this cumulant generating function,
we have derived current correlations:
cross-correlations between currents with two twisted spin in the different orbitals, cross-correlations of currents with three different spin-orbital channels, shot noises, and the Fano factor of the backscattering (nonlinear) currents.
By calculating the renormalized parameters with use of the numerical renormalization group approach,
we have investigated the exchange interaction dependence of these transport quantities.
We have discussed spin-entanglement arising in the currents through two orbitals in current cross-correlations.
It is elucidated that spin-angle dependent cross-correlation 
is induced by the exchange interactions.

Our approach can be extended 
to dots with more than two degenerate orbitals  $N>2$.
The low-energy state for the
 ferromagnetic exchange interaction is simply 
described by the $S=N/2$ Kondo state
for any orbital degeneracy $N$ and the extension can be readily obtained.
\cite{PhysRevLett.108.266401}
However, discussion on the ground state for the antiferromagnetic exchange
interaction contains more variety.
For instance,
the model for even $N$ has a critical interaction $J_{\rm c}^{}$
and the conduction electrons decouple from the dot for $J>J_{\rm c}^{}$,
similarly to the case of $N=2$ discussed in the last section.
\cite{PhysRevB.86.054421}
However, there is only one fixed point for odd $N$,
where a degenerate ground state
of the quantum dot due to the antiferromagnetic interaction yields the full-screened Kondo state.
Futher explanation of this issue is
beyond the scope of this paper.

Finally,
further study on the spin entanglement of interacting quasiparticles in the currents
may be done by the Bell's inequality or the entanglement entropy.
However, we note that not only particle pairs but also hole pairs and particle-hole pairs contribute to the spin entanglement
as seen in the generating function,
which makes it difficult to observe pure and individual entanglement of the quasiparticle pairs in the current.

\section*{Acknowldgement}
RS thanks Shiro Kawabata and Taro Wakamura for helpful discussions
and Yuya Shimazaki for inspiring discussions.
This work was partially supported by JSPS KAKENHI Grant
Nos.
JP26220711, JP26400319, JP15K05181, and JP16K17723.

\appendix

\section{Counter term}
\label{secA:CT}

The coefficients of the counter term in Eq. (\ref{eq:Lagrangean4CT}) are formally determined
by comparing both sides of Eq. (\ref{eq:Idntty-Lgrngns}) as an identity:  
\begin{eqnarray}
\xi_1^{} &=& - \Sigma_{{\rm d}m\sigma}^{\rm r} ( 0)
\, , \\
\xi_2^{} &=& z-1
\, ,\\
\xi_3^{U} &=&
z^2 \left[ U - \Gamma_{m\downarrow; m\uparrow}^{m\uparrow; m\downarrow} ( 0,0,0,0) \right]
\, ,\\
\xi_3^{W} &=& 
z^2 \left\{ W -
\frac{1}{2} \left[
\Gamma_{2\uparrow; 1\uparrow}^{1\uparrow; 2\uparrow} ( 0,0,0,0)
+ \Gamma_{2\downarrow; 1\uparrow}^{1\uparrow; 2\downarrow} ( 0,0,0,0 )
\right]
\right\} 
\, , \nonumber \\
\\
\xi_3^{J} &=&
z^2 \left[ J - \Gamma_{1 \uparrow; 2\downarrow}^{1\downarrow; 2\uparrow} ( 0,0,0,0)\right] 
\, .
\end{eqnarray}

The specific form of the counter terms as series of the renormalized interactions
for the particle-hole symmetric case $\epsilon_{\rm d}^{}= - \frac{U}{2}-W$
can be calculated order by order. 
The coefficients of the counter term up to the second order in the renormalized interactions
are obtained as
\begin{eqnarray}
	\xi_1^{} &=& 0 \, , \\
	\xi_2^{} &=&
	- \left( 3 - \frac{\pi^2}{4} \right) \widetilde{\cal I}
	 + \cdots \, ,
	\\
	\xi_3^{U} &=&
	 \pi \widetilde{\Gamma}
	 \left[
	2 \left(
	 \tilde{w} - \frac{1}{2}  \tilde{j}
	\right)^2
	-\tilde{j}^2
	\right]
	+ \cdots
	\, , \\
	\xi_3^{W} &=&
	 \pi \widetilde{\Gamma}
	\tilde{u} \tilde{j}
	+ \cdots
	\, , \\
	\xi_3^{J} &=& \pi \widetilde{\Gamma}
	\left[
	-   2 \tilde{j}^2
	-2 \tilde{u} \tilde{j}
	\right]
	+ \cdots
	\, .
\end{eqnarray}
Most of the Fermi-liquid quantities can be written
in the series of the renormalized interaction up to the second order.
Therefore, 
this obtained form of the counter term allows us to calculate the precise and specific form of Fermi liquid quantities
in terms of the renormalized parameters.

\section{Green's function for free quasiparticle}
\label{secA:GF4FQ}

The free-quasiparticle's Green's function in Eq. (\ref{eq:ActionInQP}) is given by
\begin{eqnarray}
	\widetilde{\bm g}_{{\rm d}m\sigma}^{} (t)
	= \int \frac{d\omega}{2\pi} \, \widetilde{\bm g}_{{\rm d}m\sigma}^{} ( \omega ) \, e_{}^{-i\omega t}
	 \, .
\end{eqnarray}
with the four Keldysh component in $\omega$ space,
\begin{eqnarray}
	\tilde{g}_{{\rm d}m\sigma}^{--} ( \omega )
	&=& \left[ 1- \widetilde{f}_{\rm eff}^{} ( \omega) \right] \tilde{g}_{{\rm d}m\sigma}^{\rm r} ( \omega )
	+ \widetilde{f}_{\rm eff}^{} ( \omega) \tilde{g}_{{\rm d}m\sigma}^{\rm a} ( \omega)
	\, , 
	\nonumber \\ \\ 
	\tilde{g}_{{\rm d}m\sigma}^{-+} ( \omega )
	&=& - \widetilde{f}_{\rm eff}^{} ( \omega )
	\left[ \tilde{g}_{{\rm d}m\sigma}^{\rm r} ( \omega ) - \tilde{g}_{{\rm d}m\sigma}^{\rm a} ( \omega ) \right]
\, , \\ 
	\tilde{g}_{{\rm d}m\sigma}^{+-} ( \omega )&=& 
	\left[ 1- \widetilde{f}_{\rm eff}^{} ( \omega )\right]
	\left[ \tilde{g}_{{\rm d}m\sigma}^{\rm r} ( \omega )
	- \tilde{g}_{{\rm d}m\sigma}^{\rm a} ( \omega ) \right]
	\, ,\\
	\tilde{g}_{{\rm d}m\sigma}^{++} ( \omega ) 
	&=& - \left[ 1- \widetilde{f}_{\rm eff}^{} ( \omega )\right] \tilde{g}_{{\rm d}m\sigma}^{\rm a} ( \omega )
	- \widetilde{f}_{\rm eff}^{} ( \omega ) \tilde{g}_{{\rm d}m\sigma}^{\rm r} ( \omega )
	\, .
	\nonumber \\
\end{eqnarray}
Here, the retarded and advanced Green's function of the free quasiparticles are given by
\begin{eqnarray}
\tilde{g}_{{\rm d}m\sigma}^{\rm r/a} ( \omega )
= \frac{1}{\omega - \tilde{\epsilon}_{\rm d}^{} \pm i \widetilde{\Gamma}}
\end{eqnarray}
with the full linewidth
$\widetilde{\Gamma} = \frac{1}{2} \left( \widetilde{\Gamma}_L^{} + \widetilde{\Gamma}_R^{}\right)$.
The effective Fermi distribution function for the quasiparticles is given by
\begin{eqnarray}
	\widetilde{f}_{\rm eff}^{} ( \omega )
	= \frac{\widetilde{\Gamma}_L^{} f_L^{} ( \omega ) + \widetilde{\Gamma}_R^{} f_R^{} ( \omega )}
	{\widetilde{\Gamma}_L^{} + \widetilde{\Gamma}_R^{}}
	\, ,
\end{eqnarray}
where
$f_{\alpha}^{} ( \omega ) = \left[ e_{}^{\frac{\omega - \mu_{\alpha}^{} }{T}} + 1 \right]_{}^{-1}$
is the Fermi distribution function for the electrons in the isolated left and right leads.

The quasiparticle's Green's function with counting field in the $\omega$ space in Eq. (\ref{eq:ActionwithCF}) is given by
\begin{eqnarray}
	\tilde{g}_{{\rm d}m\sigma}^{\lambda --} ( \omega ) &=&
	\frac{1}{{\cal K} ( \omega )}
	\left[ \omega - \tilde{\epsilon}_{{\rm d}}^{} - i \widetilde{\Gamma} + i \widetilde{\Gamma} \sum_{\alpha} f_{\alpha}^{} ( \omega)
	\right] \, , \\
	\tilde{g}_{{\rm d}m\sigma}^{\lambda -+} ( \omega ) &=&
	\frac{-i \widetilde{\Gamma}}{{\cal K} ( \omega)}
 	\sum_{\alpha} e^{i\lambda_{\alpha m\sigma}^{}} f_{\alpha}^{} ( \omega ) 
	\, , \\
	\tilde{g}_{{\rm d}m\sigma}^{\lambda +-} ( \omega ) &=& 
	\frac{i \widetilde{\Gamma}}{{\cal K}( \omega)} 
	\sum_{\alpha}
	e^{- i\lambda_{\alpha m\sigma}^{}} \left[ 1 - f_{\alpha}^{} ( \omega)   \right]
	\, , \\
	\tilde{g}_{{\rm d}m\sigma}^{\lambda ++} ( \omega ) &=& 
	\frac{1}{{\cal K} ( \omega )}
	\left[ - \omega + \tilde{\epsilon}_{{\rm d}}^{} + i \widetilde{\Gamma} - i \widetilde{\Gamma} \sum_{\alpha} f_{\alpha}^{} ( \omega )
	\right] \, ,
	\nonumber \\
\end{eqnarray} 
where
\begin{eqnarray}
	{\cal K} ( \omega)&=&
	\left( \omega - \tilde{\epsilon}_{\rm d}^{} \right)^2 + \widetilde{\Gamma}^2
	\nonumber \\
	&& \  + \widetilde{\Gamma}^2 \left[ e^{- i \bar{\lambda}_{m\sigma }^{} } -1 \right] 
	\left[ 1 - f_R^{} ( \omega ) \right] f_L^{} ( \omega)
\end{eqnarray}
with $\bar{\lambda}_{m \sigma}^{} = \lambda_{L m \sigma}^{} - \lambda_{R m \sigma}^{}$.

\section{Calculation of renormalized parameters}
\label{secA:CRP}

Here we briefly show how to calculate values of
$\tilde{\epsilon}_{\rm d}^{}, \widetilde{\Gamma}, \widetilde{U}, \widetilde{W}$, and $\widetilde{J}$,
using the numerical renormalization group (NRG) approach.
In this section, we fix the NRG iteration number $N$ close to the
low energy fixed point and drop the $N$ dependencies from the expressions. 
The NRG eigen energy $E$ here is not multiplied by $\Lambda^{(N-1)/2}$, where
$\Lambda$ is the logarithmic discretization parameter in NRG.

In our NRG calculation, the eigen energies are classified using a set
of quantum number $(S,Q,\Delta Q)$, 
where $S$, $Q$, and $\Delta Q$ are the total spin, the total electron
number and the difference in the number of electrons between the channels 1 and 2, respectively.
We put ${\rm Spec} H_{S,Q,\Delta Q} := \left\{ {\rm NRG\ eigen\ energies\ labeled\ with} (S,Q,\Delta Q) \right\}$.

Near the low energy fixed point corresponding to the local Fermi-liquid
state, the system is mainly described by the free quasiparticle Hamiltonian
\begin{eqnarray}
	H_{0}=
	\sum_{\kappa \eta n \sigma}{\mathcal E}^{(0)}_{\kappa,\eta,n} q^{\dagger}_{\kappa,\eta,n,\sigma} q_{\kappa,\eta,n,\sigma}^{},
\end{eqnarray}
where $q_{\kappa,\eta,n,\sigma}^{}$ annihilates a quasiparticle ($\kappa=1$) or
quasihole ($\kappa=-1$) with spin $\sigma$ and $n$ th lowest energy in the
channel 1 ($\eta=1$) or 2 ($\eta=-1$).
We consider the lowest one-quasiparticle or one-quasihole excited state $q_{\kappa,\eta,{\rm
1st},\sigma}^{\dagger} |{\rm G}\rangle$ from the ground state $|{\rm G}\rangle$.
This state is an eigen state of $H_{0}$ and has the eigen energy ${\mathcal E}^{(0)}_{\kappa,\eta,{\rm 1st}}$.
Near the low energy fixed point, we can estimate the eigen energy as
${\mathcal E}^{(0)}_{\kappa,\eta,{\rm 1st}}
=E^{(\rm 1st)}_{S_{\rm G}^{}+\frac{1}{2},Q_{\rm G}^{}+\kappa,\Delta Q_{G} +\kappa\eta}$,
where $E^{(\rm 1st)}_{S,Q,\Delta Q} :={\rm min}{\rm Spec}H_{S,Q,\Delta Q}$
and
$(S_{\rm G}^{},Q_{\rm G}^{},\Delta Q_{\rm G}^{})$ is the set of
quantum number of the ground state. (We usually have $S_{\rm G}^{}=0$.)

Since ${\mathcal E}^{(0)}_{\kappa,\eta,{\rm 1st}}$ is
determined by the poles of the Green's function for the free quasiparticles,
$\tilde{\epsilon}_{\rm d}^{}$ and
$\widetilde{\Gamma}$ can be deduced from NRG eigen energies:
\begin{widetext}
	\begin{eqnarray}
	\frac{1}{\widetilde{\Gamma}}&=&
	A_{\Lambda}\frac{2}{\pi}\frac{g_{00}^{} \left(E^{(\rm
				  1st)}_{S_{\rm G}^{}+\frac{1}{2},Q_{\rm G}^{}+1,\Delta Q_{\rm G}^{}
				  +\eta}\right)
	-
	g_{00}^{} \left(-E^{(\rm 1st)}_{S_{\rm G}^{}+\frac{1}{2},Q_{\rm G}^{}-1,\Delta Q_{\rm G}^{}
	       -\eta}\right)}
	{E^{(\rm 1st)}_{S_{\rm G}^{}+\frac{1}{2},Q_{\rm G}^{}+1,\Delta Q_{\rm G}^{} +\eta}
	+E^{(\rm 1st)}_{S_{\rm G}^{}+\frac{1}{2},Q_{\rm G}^{}-1,\Delta Q_{\rm G}^{} -\eta}}
	,\\
	\tilde{\epsilon}_{\rm d}
	&=&-A_{\Lambda}\tilde{\Gamma} \frac{2}{\pi} g_{00}^{} \left(\kappa E^{(\rm
								  1st)}_{S_{\rm G}^{} +\frac{1}{2},Q_{\rm G}^{}+\kappa,\Delta Q_{\rm G}^{} +\kappa \eta} \right),
	\end{eqnarray}
\end{widetext}
where $g_{00}^{}$ is the Green's function at site $0$ for the NRG
discretized chain of the conduction electrons and $A_{\Lambda}=\frac{1+1/\Lambda}{2(1-1/\Lambda)}\log\Lambda$
(For details, see Ref. \onlinecite{Hewson2004}.).
There is a plateau in
each of the graphs
of $\tilde{\epsilon}_{\rm d}^{}$ and
$\widetilde{\Gamma}$ against $N$ toward the low energy fixed point,
which enables us to determine the values of these renormalized parameters.

Next, we consider the following three two-quasiparticle excited states, 
\begin{eqnarray}
&& q_{\kappa,\eta,{\rm 1st},\uparrow}^{\dagger} q_{\kappa,\eta,{\rm 1st},\downarrow}^{\dagger}|{\rm G} \rangle
\, , \\
&& q_{\kappa,+1,{\rm1st},\uparrow}^{\dagger} q_{\kappa,-1,{\rm 1st}, \uparrow}^{\dagger}| {\rm G} \rangle
\, , \\
&& \frac{1}{\sqrt{2}}\left(
q_{\kappa,+1,{\rm1st},\uparrow}^{\dagger} q_{\kappa,-1,{\rm 1st},
 \downarrow}^{\dagger}-q_{\kappa,+1,{\rm1st},\downarrow}^{\dagger}q_{k,-1,{\rm
 1st}, \uparrow}^{\dagger}
 \right) | {\rm G} \rangle.
 \nonumber \\
\end{eqnarray}
The three states are eigen states of $H_{0}$,
and the eigen energies are 
$2{\mathcal E}^{(0)}_{\kappa,\eta,{\rm 1st}}$, \ ${\mathcal E}^{(0)}_{\kappa,+1,{\rm 1st}}+{\mathcal E}^{(0)}_{\kappa,-1,{\rm
1st}}$ and
${\mathcal E}^{(0)}_{\kappa,+1,{\rm 1st}}+{\mathcal E}^{(0)}_{\kappa,-1,{\rm
1st}}$,
which are shifted by $\Delta E_{1}^{}$, \ $\Delta E_{2}^{}$
 and $\Delta E_{3}^{}$, respectively,
due to the
renormalized interactions between the quasiparticles.
Near the low energy fixed point, the three energy shifts can be evaluated
using the NRG eigen energies:
\begin{eqnarray}
\Delta E_{1} &=& E_{1} - 2{\mathcal E}^{(0)}_{\kappa,\eta,{\rm
 1st}} \, ,  \\
\Delta E_{2} &=& E_{2}-({\mathcal E}^{(0)}_{\kappa,+1,{\rm 1st}}+{\mathcal E}^{(0)}_{\kappa,-1,{\rm 1st}}) 
\end{eqnarray}
and 
\begin{eqnarray}
	\Delta E_{3 }= E_{3}-({\mathcal E}^{(0)}_{\kappa,+1,{\rm 1st}}+{\mathcal
	E}^{(0)}_{\kappa,-1,{\rm 1st}}),
\end{eqnarray} 
where $E_{1}$, \ $E_{2}$, and $E_{3}$ minimize
\begin{eqnarray}
	&& \left| E-2{\mathcal E}^{(0)}_{\kappa,\eta,{\rm
	 1st}} \right|
	\ {\rm for} \ E\in{\rm Spec}H_{S_{\rm G}^{},Q_{\rm G}^{}+2\kappa,\Delta Q_{\rm G}^{}
	+2\kappa \eta} 
	\, , \nonumber\\ \\
	&& \left| E- \left( {\mathcal E}^{(0)}_{\kappa,+1,{\rm
	1st}}+{\mathcal E}^{(0)}_{\kappa,-1,{\rm 1st}} \right) \right|
	\ {\rm for}\  E\in{\rm Spec}H_{S_{\rm G}^{}+1,Q_{\rm G}^{}+2\kappa,\Delta
	Q_{\rm G}^{}} \, ,
	\nonumber \\
\end{eqnarray}
and
\begin{eqnarray}
	&& \left| E- \left({\mathcal E}^{(0)}_{\kappa,+1,{\rm
	1st}}+{\mathcal E}^{(0)}_{\kappa,-1,{\rm 1st}} \right) \right| \ {\rm for} \ E\in{\rm Spec}H_{S_{\rm G}^{},Q_{\rm G}^{}+2\kappa,\Delta
	Q_{\rm G}^{}} \, ,
	\nonumber \\
\end{eqnarray}
respectively.
Since the effects of the renormalized interactions on the low-lying many-particle excitations tend to zero toward 
the low energy fixed point, the three energy shifts can be
calculated
using the ordinarily first order perturbation expansion in
the renormalized
interactions, 
which gives us the following three formulas for calculating the renormalized
interaction parameters,
\begin{eqnarray}
	\widetilde{U} &=&
	\frac{\Delta E_{1}}{|C^{(0)}_{\kappa,\eta,{\rm 1st}}|^{4}},\\
	\widetilde{W} &=&
	\frac{3\Delta E_{2}+\Delta E_{3}}{4|C^{(0)}_{\kappa,+1,{\rm
	1st}}|^{2}|C^{(0)}_{\kappa,-1,{\rm 1st}}|^{2}},\\
	2\widetilde{J}&=&
	\frac{\Delta E_{2}-\Delta E_{3}}{|C^{(0)}_{\kappa,+1,{\rm
	1st}}|^{2}|C^{(0)}_{\kappa,-1,{\rm 1st}}|^{2}}.
\end{eqnarray}

Here, $C^{(0)}_{\kappa,\eta,{\rm 1st}}$ is the overlap integral between the
orbital of the dot coupled with the channel $\eta$ and the free quasiparticle
state labeled with $(\kappa,\eta,{\rm 1st})$.
Toward the low energy fixed point,
we usually have a plateau in
each of the graphs of 
$\widetilde{U}, \widetilde{W}$, and $\widetilde{J}$
against $N$, which enable us to evaluate
the renormalized interaction parameters.


%

\end{document}